\documentclass[aps,prd,reprint,preprintnumbers,superscriptaddress,nofootinbib,longbibliography,floatfix]{revtex4-2}
\usepackage[T1]{fontenc}
\usepackage{amsmath,amssymb,amsthm}
\usepackage{graphicx}
\usepackage{xcolor}
\usepackage{booktabs}
\usepackage{enumitem}
\usepackage{microtype}
\usepackage[colorlinks,linkcolor={red!50!black},citecolor={blue!50!black},urlcolor={blue!80!black}]{hyperref}
\hypersetup{pdftitle={Evaluating generative samplers for lattice field theory by scale}, pdfauthor={Satoshi Iso, Lingxiao Wang}}
\graphicspath{{figures/}}
\newtheorem{proposition}{Proposition}
\newcommand{\todo}[1]{}
\newcommand{\Eq}{\mathbb{E}_q}
\newcommand{\Ep}{\mathbb{E}_p}
\newcommand{\ESS}{\mathrm{ESS}}
\newcommand{\TV}{\mathrm{TV}}
\newcommand{\kh}{\hat k}
\newcommand{\nref}{n_{\mathrm{ref}}}
\newcommand{\chat}{\hat c}

\begin{document}

\title{ Evaluating generative samplers for lattice field theory by scale }

\preprint{RIKEN-iTHEMS-Report-26}

\author{Satoshi Iso}
\email{satoshi.iso@riken.jp}
\affiliation{RIKEN Center for Interdisciplinary Theoretical and Mathematical Sciences (iTHEMS), Wako, Saitama 351-0198, Japan}
\affiliation{KEK Theory Center, Institute of Particle and Nuclear Studies, KEK, Oho 1-1, Tsukuba, Ibaraki 305-0801, Japan}
\author{Lingxiao Wang}
\email{lingxiao.wang@riken.jp}
\affiliation{RIKEN Center for Interdisciplinary Theoretical and Mathematical Sciences (iTHEMS), Wako, Saitama 351-0198, Japan}
\affiliation{Institute for Physics of Intelligence, Graduate School of Science, The University of Tokyo, Bunkyo-ku, Tokyo 113-0033, Japan}

\date{\today}

\begin{abstract}
Generative samplers for lattice field theory are commonly evaluated by their effective sample size (ESS). We show that a small global ESS need not imply inaccurate long-distance observables: errors spread over many short-distance modes can dominate this diagnostic.  We introduce a scale-resolved diagnostic based on the overlap between coarse-grained model and target distributions. At the lattice cutoff, this overlap determines the asymptotic ESS fraction; lower cutoffs isolate long-wavelength errors.  In the exactly solvable free theory, two samplers with comparable ESS yield a $32\%$ bias and no bias in the susceptibility, respectively.  In two-dimensional $\phi^4$ theory near criticality, trained diffusion samplers show the same separation of scales: their path-weight ESS fraction falls rapidly over the volumes studied, while low moments of the magnetization remain close to their target values. Varying the reference parameters substantially changes infrared observables, with little effect on the log-weight variance. Given the target magnetization distribution, a marginal Metropolis test maintains an acceptance rate of order one in our tests. Combined with exact local updates, it reduces the remaining mismatch, although residual bias remains at finite relaxation length. In the free field, uniformly bounded relative errors in the proposal mode variances give a volume-independent bound on the number of updates needed for a prescribed accuracy of the chain's stationary mode variances at fixed correlation length. The target marginal must be supplied or computed separately; estimating it from the samplers' own weights can inherit the poor global overlap. 
\end{abstract}

\maketitle

\renewcommand{\topfraction}{0.9}\renewcommand{\dbltopfraction}{0.9}\renewcommand{\bottomfraction}{0.6}
\renewcommand{\textfraction}{0.08}\renewcommand{\floatpagefraction}{0.75}\renewcommand{\dblfloatpagefraction}{0.75}
\setcounter{topnumber}{3}\setcounter{dbltopnumber}{3}

\section{Introduction}\label{sec:intro}

Generative models are increasingly used to sample field configurations in lattice field theory; for reviews of machine-learning methods in lattice QCD, see Refs.~\cite{Cranmer:2023,Aarts:2025b,Tomiya:2025quf}. Examples include normalizing flows~\cite{Albergo:2019,Kanwar:2020,DelDebbio:2021,Abbott:2022,Cheng:2025} and diffusion models, which are closely related to stochastic quantization~\cite{Wang:2023,Fukushima:2024}. Diffusion models have been applied to scalar field theories and sigma models~\cite{Vega:2025,Tan:2026,Grinwald:2026,Schubring:2026} and to gauge theories~\cite{Zhu:2025,Aarts:2026,Alharazin:2026,Komijani:2026}, also with dynamical fermions~\cite{Vega:2026,Tomiya:2026b}; they have been extended to theories with a sign problem~\cite{Aarts:2025} and trained without reference ensembles~\cite{Tomiya:2026,Grinwald:2026}. Other samplers are based on stochastic quantization formulated as optimal control (OSQ)~\cite{Wang:OSQ}, on stochastic path sampling~\cite{Chen:SPS} or on non-equilibrium transport with a learned drift~\cite{Albergo:2024}. Because the model distribution $q$ differs from the target distribution $p=e^{-S}/Z$, asymptotically exact results require a correction: reweighting with the importance weights $w=p/q$ ~\cite{Nicoli:2020}, an independence Metropolis--Hastings (MH) test ~\cite{Albergo:2019},  or  Metropolis-adjusted Langevin steps~\cite{Zhu:2025,Aarts:2026}  and  MH tests~\cite{Tomiya:2026} during generation.

The central obstacle to this program is the scaling with the number of degrees of freedom. For flow-based samplers of $\phi^4$ theory, the cost of training a model to a given sampling efficiency grows extremely rapidly as the continuum limit is approached~\cite{DelDebbio:2021}, and flow-based trivializing maps with a fixed architecture reduce the autocorrelation times of hybrid Monte Carlo (HMC) without changing their scaling toward the continuum~\cite{Albandea:2023}. This poor scaling has been attributed to the multiscale structure of the problem near critical points, which networks of moderate size do not capture efficiently~\cite{Komijani:2023}; in diffusion models the same structure makes the score ill-conditioned, so that the number of time steps of the generating process can grow, a difficulty mitigated by factorizing the distribution across scales~\cite{Guth:2022}. The dependence on the volume is even more severe: the quality of a fixed model built from local transformations degrades exponentially under volume transfer~\cite{Abbott:2022}, and for local theories a single global independence proposal can become impractical at realistic box sizes~\cite{Finkenrath:2024}. The scalability of flow-based sampling must therefore be assessed experimentally~\cite{Abbott:2022}; see also Ref.~\cite{Cranmer:2023}. These studies suggest that, for a fixed local mismatch and finite correlation length, the ESS fraction and the acceptance rate of a global independence test can decrease as $e^{-cV}$, with a model-dependent constant $c$. The volume dependence of the ESS of the OSQ sampler reported in Ref.~\cite{Wang:OSQ} is consistent with this behavior qualitatively. The usual solution has been to reduce $c$ through more expressive architectures and improved training.

The exponential volume dependence has a simple origin. If the model deviates slightly from the target in each correlation volume, $\log w$ is a sum of many weakly correlated contributions, and its variance grows approximately in proportion to $V$. For independent subvolumes, the population ESS fraction factorizes~\cite{Abbott:2022}. The useful sample size, however, depends on the observable~\cite{Abbott:2022}, and flow-based variance reduction for specific observables has been found to be approximately independent of the volume~\cite{Abbott:2026}; a flow that encodes the generating functional extends it to arbitrary correlation functions~\cite{Abbott:2026b}.

Other proposals modify the structure of the sampler. Stochastic normalizing flows interleave flow layers with out-of-equilibrium Monte Carlo updates and inherit the favorable scaling of non-equilibrium Monte Carlo with the number of degrees of freedom~\cite{Caselle:2022,Bulgarelli:2024,Bonanno:2026}, also when the intermediate distributions are those of a trained energy-based diffusion model~\cite{Kanwar:2026}, closely related to sequential Monte Carlo along the generating process~\cite{Sexty:2026}. Continuous normalizing flows improve the sampling efficiency, in particular on larger lattices~\cite{Gerdes:2023}, and learned gradient flows reduce the number of trainable parameters by orders of magnitude~\cite{Bacchio:2023}. Localized updates restrict the generative proposal to subdomains~\cite{Finkenrath:2022,Finkenrath:2024}, and renormalization-group-inspired constructions treat different length scales separately~\cite{Li:2018,Marchand:2022,Abbott:2024,Bauer:2024,Singha:2025,Singha:2026}, also in diffusion and flow-matching models, which generate the field scale by scale~\cite{Guth:2022,Schubring:2026} or along a renormalization-group flow~\cite{Cotler:2023,Masuki:2025,Masuki:2026}. Physics-informed kernels (PIKs) build the generative map layer by layer from renormalization-group flows, so that learning reduces to solving a set of independent linear differential equations for the kernels of the transformation~\cite{Kapust:2025}; the construction has been extended to complex distributions with a sign problem~\cite{Ihssen:2026}. Inverse blocking avoids the global test altogether: a coarse ensemble matched by a renormalization-group blocking supplies the long-distance modes, a conditional flow reconstructs the short-distance degrees of freedom, and rethermalization with the fine action reduces the residual mismatch~\cite{Hasenfratz:2026,Hasenfratz:2026b}. Short-distance mismatches then relax rapidly, whereas a mismatch in the relevant thermal direction relaxes much more slowly~\cite{Hasenfratz:2026}, as expected for an error that moves the ensemble away from the critical surface. Similarly, the residual bias of a diffusion-model sampler for $\phi^4$ theory has been found to be concentrated in the zero mode~\cite{Tan:2026}.

\begin{samepage}
These observations motivate two quantitative questions:
\begin{enumerate}[label=(\arabic*),leftmargin=2em]
\item Which part of the mismatch determines the ESS, and which part affects physical observables?
\item Given the target distribution of selected infrared variables, how can it be used to reduce errors in full-configuration proposals? 
\end{enumerate}
\end{samepage}
 We develop a framework for evaluating sampler errors by scale, the main subject of this paper, and examine both questions in the free field.  The asymptotic ESS fraction is $e^{-D_2(p\|q)}$, where $D_2$ is the R\'enyi divergence of order two. Removing short-distance modes gives a monotone overlap flow $D_2(\Lambda)$: its value at the lattice cutoff determines the global ESS, while smaller cutoffs isolate errors at longer wavelengths. We construct two samplers with similar global ESS: a mass error biases the susceptibility by $32\%$, whereas a short-distance lattice artifact leaves it unbiased. The distinction also explains why weighting only the modes on which an observable depends avoids noise from other modes, in the spirit of the local factorization of Ref.~\cite{Ce:2016}.

For the second question, we derive the mean acceptance rate of an independence test and show that a test on a few collective variables (low-dimensional functions of the configuration such as the magnetization), which we call the marginal test, accepts at least as often as the full test  at their respective stationary distributions. Its acceptance stays bounded away from zero if the overlap of the chosen marginal distributions does so as the volume grows. We then use exact local updates to relax the errors in the remaining modes. In the free field, the resulting composite chain reaches a given accuracy of the mode variances with a number of updates independent of volume at fixed correlation length and uniformly bounded relative errors in the proposal mode variances.  Thus, once the infrared input is supplied, poor global overlap need not imply exponentially costly relaxation to a prescribed accuracy. 

We test this picture with trained OSQ samplers for two-dimensional $\phi^4$ theory near criticality.  With reference parameters fixed from the benchmark, they  reproduce $\langle|M|\rangle$ and $\langle M^2\rangle$ to a few percent while their path-weight ESS fractions fall below $10^{-3}$ at $L=64$; the test on $M$ still accepts about $0.8$ of the proposals. Changing the reference parameters substantially changes the infrared observables while barely changing the log-weight variance. In the composite chain, about ten HMC trajectories remove most of the ultraviolet mismatch at every volume studied.  The order of the two corrections matters when the proposal's ultraviolet is far from the target: its relaxation shifts the tested variables, so the ultraviolet should be relaxed before the infrared is corrected, or with the tested variables held fixed.  The remaining limitations concern the untested low modes, their coupling to the tested variables, and errors in the estimated marginal densities.  We also estimate the target distribution of $M$ from the action by integrating the mean force at fixed $M$ at $L=32$ and $64$, whereas the samplers' own weights give it reliably only at $L=16$ (Appendix~\ref{app:pphi}). 

The marginal test alone has the correct stationary distribution for the tested variables when their density ratio is known; at fixed values of these variables it retains the model's conditional distribution. The composite chain is generally biased at finite relaxation length. We therefore distinguish the cost set by full microscopic overlap from that set by a few infrared variables and the local relaxation time. The practical question is whether the relevant infrared errors can be identified and corrected using a number of variables that does not grow with volume, and whether their target distribution can be obtained at controlled cost as the volume grows.  This approach is an alternative to inverse blocking~\cite{Hasenfratz:2026,Hasenfratz:2026b}: instead of a matched coarse ensemble, which supplies the joint distribution of the long-distance modes implicitly, the marginal test targets the joint distribution of a few chosen variables, preferably those that local relaxation repairs most slowly, and needs it as input. The two could be combined, but such a hybrid has not been tested (Sec.~\ref{sec:discussion}). 

The paper is organized as follows. Sections~\ref{sec:ess} and~\ref{sec:IR} give the general framework: Sec.~\ref{sec:ess} relates the ESS to the R\'enyi divergence and introduces the overlap flow, and Sec.~\ref{sec:IR} constructs the marginal test and the composite chain. Sections~\ref{sec:free} and~\ref{sec:phi4} test both parts in parallel, exactly in the free field and numerically in two-dimensional $\phi^4$ theory near its critical point. Section~\ref{sec:discussion} discusses what evaluating samplers by scale implies for their assessment and design, and Sec.~\ref{sec:concl} concludes. Derivations are collected in  Appendix~\ref{app:deriv},  Appendix~\ref{app:fixedxi} studies the composite chain at fixed correlation length, and Appendix~\ref{app:pphi} examines how to estimate and use the target marginal.

\section{Effective sample size and coarse graining}\label{sec:ess}

We first relate the effective sample size to the R\'enyi divergence and define a coarse-grained version of it. The overlap identities apply to normalized nonnegative target and model distributions; the scaling statements require the assumptions specified below.

\subsection{R\'enyi divergence}
Let $w=p/q$, with $q>0$ wherever $p>0$, so $\Eq[w]=1$. For independent samples $\phi_i\sim q$ and finite $\Eq[w^2]$, the usual estimate $\widehat\ESS=(\sum_iw_i)^2/\sum_iw_i^2$ satisfies
\begin{equation}\label{eq:ESS-D2}
\begin{gathered}
\frac{\widehat{\ESS}}{N}\ \xrightarrow{N\to\infty}\ \frac{1}{\Eq[w^2]}=e^{-D_2(p\|q)},\\
D_2(p\|q)=\log\int d\phi\,\frac{p^2}{q}.
\end{gathered}
\end{equation}
The ratio follows from $\Eq[w]=1$ and $\Eq[w^2]=\int p^2/q$. The normalization of $p$ or $q$ need not be known: with $\ell=\log w$ given up to an additive constant,
\begin{equation}\label{eq:D2-unnorm}
D_2=\log\Eq\big[e^{2\ell}\big]-2\log\Eq\big[e^{\ell}\big],
\end{equation}
which does not change when a constant is added to $\ell$. If $\ell$ is Gaussian with variance $\sigma^2$, Eq.~\eqref{eq:D2-unnorm} gives $D_2=\sigma^2$. Thus $\ESS/N\simeq e^{-\mathrm{Var}(\log w)}$ is exact for Gaussian log weights; otherwise the exponential moments in Eq.~\eqref{eq:D2-unnorm}, especially their sensitivity to the upper tail, must be considered.

For translation-invariant targets at fixed finite correlation length, a weak local mismatch gives an extensive $D_2$. A sampler defines an effective action $-\log q$; if it differs from $S$ by a local term, $q=e^{-S-cA}/Z(c)$ with $A=\sum_xO(x)$, then, provided $Z(c)$ is finite in a neighborhood of $c=0$ containing $\pm c$ (Appendix~\ref{app:gibbs}),
\begin{equation}\label{eq:gibbs}
\begin{gathered}
D_2=\log Z(c)+\log Z(-c)-2\log Z(0)\simeq c^2\,V\chi_O,\\
\chi_O=\sum_x\langle O(x)O(0)\rangle_{\rm conn}.
\end{gathered}
\end{equation}
The next term is $c^4\kappa_4(A)/12$, with $\kappa_4$ the fourth cumulant. For several perturbations $\sum_ic_iA_i$ the leading term is $\sum_{ij}c_ic_j\langle A_iA_j\rangle_{\rm conn}$, which is the Fisher metric of the Gibbs family. The negative logarithm of the population ESS fraction is thus a second difference of the free energy and grows like $V$ when $\chi_O$ and the higher cumulant densities remain finite. This need not imply inaccurate long-distance observables: $D_2$ simply counts how many regions carry the same small error.

\subsection{Overlap flow}\label{sec:flow}
Let $R_\Lambda$ keep the Fourier modes with lattice momentum $\kh\le\Lambda$, where $\kh^2=\sum_\mu4\sin^2(k_\mu/2)$, and define $D_2(\Lambda)=D_2(R_\Lambda p\|R_\Lambda q)$. This compares the Wilsonian effective actions after removing modes above $\Lambda$. Coarse graining averages over the discarded fluctuations: $Rp/Rq=\Eq[w\,|\,R(U)]$. Jensen's inequality then gives $\int Rq\,(Rp/Rq)^2\le\Eq[w^2]$ (Appendix~\ref{app:dpi}; general Markov kernels, including other block-spin maps, are covered in Ref.~\cite{vanErven:2014}). Lowering $\Lambda$ therefore decreases $D_2(\Lambda)$ monotonically, from the global value at the lattice cutoff to the zero-mode value at $\Lambda=0$. The proof needs neither Gaussianity nor translation invariance and holds with interactions.

\subsection{Relevant operators and sectors}\label{sec:sectors}
For a weak mismatch $c\sum_xO(x)$ with scaling dimension $\Delta$ in the regime $a\ll\xi\ll L$, Eq.~\eqref{eq:gibbs} gives $D_2\simeq c^2V\chi_O$, and power counting gives $\chi_O$ scaling as $\xi^{d-2\Delta}$ (IR dominated) for $\Delta<d/2$, as $\log(\xi/a)$ for $\Delta=d/2$, and as $a^{d-2\Delta}$ (UV dominated) for $\Delta>d/2$. This boundary differs from the RG boundary $y=d-\Delta=0$: long-distance scaling is especially sensitive to relevant couplings ($y>0$), marginal directions, the zero mode and sector weights, while irrelevant couplings generally give suppressed corrections. Operators with $\Delta>d/2$ contribute to $D_2$ mainly at the cutoff scale in this approximation. The mass mismatch ($\Delta=0$) and the lattice artifact ($\Delta=4$) of the two-dimensional free field (Sec.~\ref{sec:free}) are examples of the two extremes. If the configuration space splits into sectors $a$ (vacua, topological charge),
\begin{equation}\label{eq:sector}
e^{D_2(p\|q)}=\sum_a\frac{p_a^2}{q_a}\,e^{D_2(p(\cdot|a)\|q(\cdot|a))},
\end{equation}
so a missing sector ($q_a=0$, $p_a>0$), as in a mode-collapsed sampler~\cite{Nicoli:2023}, makes $D_2$ infinite. Reweighting can correct wrong sector probabilities only if the proposal reaches every sector populated by the target.

\section{Metropolis test on collective variables}\label{sec:IR}

Exact results for all observables require that the short-distance errors be removed as well. The inverse population ESS fraction of global reweighting, $N/\ESS\simeq e^{D_2}$, grows exponentially with $V$ when $D_2$ is extensive (Sec.~\ref{sec:ess}). For an observable depending only on a few variables, reweighting by their marginal density ratio instead incurs the ESS loss of that marginal; Sec.~\ref{sec:estimators} demonstrates this in the free field. We use the same kind of low-dimensional information in a Metropolis test on a few collective variables and let exact local updates relax the remaining errors (Sec.~\ref{sec:fix}).

\subsection{Acceptance rate}
In an independence MH chain with proposal $y\sim q$ and current state $x\sim p$, the mean acceptance rate at stationarity is exactly (Appendix~\ref{app:acc})
\begin{equation}\label{eq:alpha-TV}
\begin{aligned}
\alpha&=\int dx\,dy\,\min\big[p(x)q(y),\,q(x)p(y)\big]\\
&=1-\TV(p\otimes q,\ q\otimes p)\ \le\ e^{-D_{1/2}(p\|q)},
\end{aligned}
\end{equation}
where $\TV(f,g)=\frac12\int|f-g|$ is the total variation distance between probability densities $f$ and $g$, and $D_{1/2}=-2\log\int\sqrt{pq}$. Like $D_2$, $D_{1/2}$ is extensive for a fixed local mismatch at finite correlation length. For independence proposals with a positive divergence per unit volume, this gives an exponentially small upper bound on acceptance. For Gaussian $\log w$ one finds $\alpha=2\Phi_{\rm N}(-\sigma/\sqrt2)\le e^{-\sigma^2/4}=e^{-D_{1/2}}$, and $\alpha\simeq(2/\sigma\sqrt\pi)\,e^{-\sigma^2/4}$ for $\sigma\gg1$, where $\Phi_{\rm N}$ is the standard normal cumulative distribution function.

\subsection{Marginal test}
Take a few collective variables $\Phi(U)$ (magnetization $M$, topological charge $Q$, \dots) with marginals $p_\Phi$, $q_\Phi$. Propose $U'\sim q$ and accept with
\begin{equation}\label{eq:margtest}
a(\Phi\to\Phi')=\min\left[1,\frac{p_\Phi(\Phi')\,q_\Phi(\Phi)}{p_\Phi(\Phi)\,q_\Phi(\Phi')}\right].
\end{equation}
The model density $q(U)$ is never needed, so the test applies also to models whose density is not known in closed form.

\begin{proposition}[Marginal acceptance]\label{prop:marg}
With the current collective variables distributed as $p_\Phi$, the mean acceptance rate of \eqref{eq:margtest} is
\[
\alpha_\Phi=1-\TV(p_\Phi\otimes q_\Phi,\,q_\Phi\otimes p_\Phi)\ \ge\ \alpha.
\]
It depends only on the distributions of $\Phi$. It stays bounded away from zero if the overlap of the appropriately rescaled marginal distributions does so as the volume grows.
\end{proposition}
The inequality holds because mapping $(x,y)\mapsto(\Phi(x),\Phi(y))$ can only reduce a total variation distance (Appendix~\ref{app:acc}). For example, if $M$ is approximately Gaussian with zero mean under both distributions, $\alpha_M$ depends only on the variance ratio $\langle M^2\rangle_q/\langle M^2\rangle_p$. Near the critical point the distribution of $M$ is not Gaussian, and $\alpha_\Phi$ has to be computed from it. A marginal test can also become inefficient if its own target and proposal distributions lose overlap.

\subsection{Composite Markov chain}\label{sec:fix}
With the exact marginal density ratio and proposal support covering the target marginal, the marginal test alone has stationary distribution (Appendix~\ref{app:stat})
\begin{equation}\label{eq:stationary}
\pi(U)=p_\Phi(\Phi)\,q(U|\Phi).
\end{equation}
 The collective variables have their target distribution; the remaining variables retain the model's conditional distribution. If $p_\Phi$ is known, observables of $\Phi$ alone can already be evaluated directly. The additional role of a configuration chain is to sample the other modes and their correlations with $\Phi$.  We therefore follow each test by $\nref$ steps of an exact local kernel $K_{\rm loc}$ (HMC) with $pK_{\rm loc}=p$. The composite chain is generally \emph{not} exact at finite $\nref$. When local relaxation is exponential, its residual bias is expected to decay like $e^{-\nref/\tau}$, with $\tau$ set by the slow modes coupled to the proposal error. For a zero-mode test in the free field, the tested mode decouples and the remaining modes relax in parallel; the mode-variance bound of Sec.~\ref{sec:exact} is set by the slowest untested mode, not their number. Interactions can couple this relaxation back to the tested variables (Sec.~\ref{sec:phi4chain}). For $L\gg\xi$ we expect $\tau\sim\xi^z$, independent of $L$. In the free field this can be proved; we do so in Sec.~\ref{sec:exact}. For a test on $\Phi=M$ near criticality, where $\xi$ is comparable to $L$, a free-field estimate identifies the lowest nonzero Fourier modes as the slowest untested modes, and $\tau$ grows with $L$ as expected for the local dynamics considered here; Sec.~\ref{sec:phi4chain} studies this case in $\phi^4$ theory. In this estimate, when $\xi$ grows in proportion to $L$, the shells whose relaxation time is within a factor of a few of the slowest form a fixed set, independent of $L$, so they can be included in $\Phi$ , at the price of estimating a joint density of more variables (Sec.~\ref{sec:discussion}) . In the decoupled free-field picture, this shortens $\tau$ by a constant factor but does not change its growth with $\xi$.

The test needs only the ratio $r(\Phi)=p_\Phi(\Phi)/q_\Phi(\Phi)$ of two densities of a few variables, up to a constant.  The model marginal $q_\Phi$ can be estimated from a pool of independent proposals.  The target marginal $p_\Phi$ is the actual input of the method.  It may be available analytically, from a validated finite-size scaling form, or through the constraint effective potential   $-V^{-1}\log p_\Phi$ (up to a constant) ~\cite{ORaifeartaigh:1986}, which can itself be computed in simulations at fixed $\Phi$; for the magnetization, $-\partial_M\log p_M=\langle\sum_x\partial S/\partial\phi_x\rangle_M$, and such simulations only need to relax the modes outside $\Phi$.  For $\Phi=M$ they can be started from generator samples shifted uniformly to the given value of $M$; in $\phi^4$ theory at $L=32$ and $64$ the resulting estimate $\hat p_M$ reproduces the benchmark distribution of $M$ (Appendix~\ref{app:pphi}). Drawing $M$ from $\hat p_M$ and relaxing a shifted generator sample at fixed $M$ then gives proposals whose distribution of $M$ is $\hat p_M$, so that no estimate of $q_M$ is needed; the error of $\hat p_M$ and the part of the conditional error that the relaxation does not remove remain.   A coarse-grained or auxiliary ensemble can also be used, with its own systematic error. For a discrete variable such as the topological charge there is no mean force, and simulations at fixed charge do not directly give the relative weights of the sectors; these require additional computations, for example metadynamics along a smoothed charge, which reconstructs $-\log p(Q)$~\cite{Laio:2016}.  Instead of estimating $p_\Phi$ and $q_\Phi$ separately, $\log r$ can also be fitted directly, for example by logistic regression between model and target samples of $\Phi$, which avoids taking the ratio of two noisy estimates in the tails.  The identity $r=\Eq[w\,|\,\Phi]$ is exact (Sec.~\ref{sec:flow}), but estimating this conditional average can inherit large weight fluctuations from the untested modes. The model's own weights therefore do not generally provide the infrared marginal at low cost (Appendix~\ref{app:pphi}); the conditional fluctuations vanish only when the weight depends on $\Phi$ alone.  An error in $r$ changes only the distribution of $\Phi$ enforced by the test; it is a low-dimensional systematic error that has to be quoted (Sec.~\ref{sec:phi4chain}).

\section{Free field theory}\label{sec:free}

For the Gaussian target and proposal below, the overlap quantities of Sec.~\ref{sec:ess} and the composite chain of Sec.~\ref{sec:fix} can be computed exactly, mode by mode. This section is the exactly solvable counterpart of Sec.~\ref{sec:phi4}: Secs.~\ref{sec:mismatch} and~\ref{sec:estimators} address the first question of the introduction, and Sec.~\ref{sec:exact} quantifies the relaxation cost at prescribed accuracy for the second.

\subsection{Mode decomposition}\label{sec:modes}
Take a free field on an $L\times L$ lattice with $V=L^2$. We use real orthonormal modes $x_k$, so $x_0=\sqrt V M$ with $M=V^{-1}\sum_x\phi_x$, and write $G_k=\langle x_k^2\rangle$ and $\chi=V\langle M^2\rangle=G_0$. Each normal mode is Gaussian with inverse variance $\lambda_k=\kh^2+m^2$, where $\kh$ is defined in Sec.~\ref{sec:flow} and $m=1/\xi$. The model inverse variance is $\lambda_k(1+\varepsilon_k)$. For $|\varepsilon_k|<1$, integrating each mode gives (Appendix~\ref{app:mode})
\begin{equation}\label{eq:D2free}
D_2=-\frac12\sum_k\log(1-\varepsilon_k^2)\simeq\frac12\sum_k\varepsilon_k^2.
\end{equation}
For small errors each mode contributes $\varepsilon_k^2/2$ regardless of wavelength. We compare a mass error, strongest at long wavelengths, with a lattice artifact, strongest at short wavelengths:
\begin{equation}\label{eq:profiles}
\begin{aligned}
&\text{relevant (mass):} && \varepsilon_k=\frac{\chat\,m^2}{\kh^2+m^2},\\
&\text{irrelevant (lattice artifact):} && \varepsilon_k=\frac{c_6\,\kh^4}{\kh^2+m^2},
\end{aligned}
\end{equation}
with $\chat=\delta m^2\xi^2$. The relevant one is largest at $k=0$; the irrelevant one vanishes at $k=0$ and is $O(c_6)$ at the cutoff.

\subsection{Relevant and irrelevant mismatches}\label{sec:mismatch}
\paragraph{Relevant mismatch.} For small $\chat$ in the continuum, large-box approximation (Appendix~\ref{app:corrvol}),
\begin{equation}\label{eq:corrvol}
D_2\simeq\frac{1}{8\pi}\left(\frac{L}{\xi}\right)^2\chat^2.
\end{equation}
The negative logarithm of the ESS fraction counts correlation volumes $N_{\rm corr}=(L/\xi)^2$. The exact lattice sums follow this line within about 13\% (Fig.~\ref{fig:collapse} in Appendix~\ref{app:corrvol}). Keeping the population ESS fraction fixed needs $\chat\lesssim N_{\rm corr}^{-1/2}$: the model must become more accurate as $V$ grows.

\paragraph{Irrelevant mismatch.} Every mode near the cutoff contributes $O(c_6^2)$, so $D_2$ is proportional to the number of sites. At fixed physical volume, $D_2\propto V_{\rm phys}/a^2\to\infty$ as $a\to0$, while at a fixed physical momentum $p$ the mismatch $\varepsilon\simeq c_6(ap)^2$ goes to zero. The fixed-momentum observables have the same continuum limit, although the global ESS fraction vanishes.\footnote{For a fixed pair of infinite-dimensional Gaussian measures, equivalence or mutual singularity is governed by the Feldman--H\'ajek dichotomy \cite{Feldman:1958,Hajek:1958}; for small $\varepsilon_k$ the measures are equivalent only if $\sum_k\varepsilon_k^2<\infty$.} For this irrelevant mismatch, loss of global overlap therefore coexists with accurate long-distance observables.

\begin{figure}[t]
\centering
\includegraphics[width=\columnwidth]{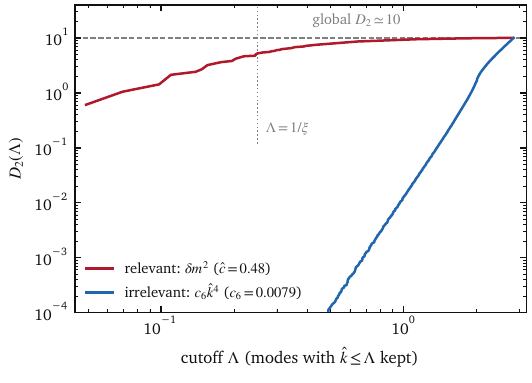}
\caption{Comparable global ESS, different susceptibility bias (exact lattice sums, $L=128$, $\xi=4$). The global $D_2$ values are $10.09$ and $9.99$, both giving $\ESS/N\simeq4\times10^{-5}$. Reading from right to left removes short-distance modes. Red: the mass error saturates at $\Lambda\sim1/\xi$ and survives in the infrared, biasing the susceptibility by $-32\%$. Blue: the lattice artifact falls roughly as $\Lambda^6$ for $m\ll\Lambda\ll1$ and vanishes at the zero mode, leaving the susceptibility exact.}
\label{fig:flow}
\end{figure}

Removing short-distance modes distinguishes the two mismatches (Appendix~\ref{app:corrvol}):
\begin{equation}\label{eq:flowrel}
\begin{aligned}
\text{relevant:}\quad & D_2(\Lambda)\simeq D_2^{\rm glob}\frac{\Lambda^2}{m^2+\Lambda^2},\quad \text{IR end }\simeq\chat^2/2,\\
\text{irrelevant:}\quad & D_2(\Lambda)\propto\Lambda^6\ \ (m\ll\Lambda\ll1),\quad \text{IR end }=0.
\end{aligned}
\end{equation}

Figure~\ref{fig:flow} shows why the scale matters. The susceptibility measures the zero-mode variance: a mass error changes it, whereas the lattice artifact vanishes at $k=0$. The global overlap combines both long- and short-wavelength errors and obscures this distinction.

\paragraph{Ultraviolet dominance.} Replacing the sum \eqref{eq:D2free} by an integral over scales,
\begin{equation}\label{eq:scaleintegral}
-\log\frac{\ESS}{N}\simeq\int^{\Lambda_{\rm UV}}d\log\Lambda\;\underbrace{\frac{dN_{\rm modes}}{d\log\Lambda}}_{\propto\,V\Lambda^d}\;\frac{\varepsilon(\Lambda)^2}{2}.
\end{equation}
The number of modes per logarithmic interval grows as $\Lambda^d$. Short wavelengths dominate when their relative error falls more slowly than $\Lambda^{-d/2}$. A pure mass error falls as $\Lambda^{-2}$ in two dimensions and instead gives the correlation-volume law. Thus a persistent error near the cutoff can dominate the ESS as $a\to0$, even when the long-wavelength error vanishes.

\subsection{Reweighting}\label{sec:estimators}

Let $O$ depend only on a set $A$ of modes, and $w_A=\prod_{k\in A}w_k$, where $w_k=p_k(x_k)/q_k(x_k)$ is the normalized single-mode weight. In the free field, provided the required reweighted moments are finite (proof in Appendix~\ref{app:three}):
\begin{proposition}[Three estimators]\label{prop:three}
(i) Local reweighting $\Eq[w_AO]/\Eq[w_A]$ is exact, with ESS fraction $e^{-D_2(A)}$, independent of $V$ when both $A$ and its mode mismatches are fixed.
(ii) Without weights, the relative bias of $|x_k|^2$ is $-\varepsilon_k/(1+\varepsilon_k)$: a volume-independent response at fixed $\varepsilon_k$.
(iii) Global reweighting multiplies the asymptotic variance by exactly $e^{D_2-D_2(A)}$. The weights of the modes that $O$ does not see correct nothing and only add noise.
\end{proposition}

\begin{table*}[t]
\centering\small
\begin{tabular}{llccc}
\toprule
mismatch & observable & no weights & local weights & global weights\\
\midrule
relevant ($\chat=0.77$) & $\chi$ & $9.05(7)$ [$-43\%$] & $\mathbf{16.05(57)}$ & $15.27(1.57)$\\
irrelevant ($c_6=0.013$) & $\chi$ & $\mathbf{16.07(14)}$ [$\approx0$] & $16.07(14)$ & $16.30(1.70)$\\
\bottomrule
\end{tabular}
\caption{Free field, $L=64$, $\xi=4$, both mismatches tuned to $D_2\simeq7$ ($7.05$ and $6.77$), $N=2\times10^4$ samples per run, 40 runs. Entries: mean over the runs and, in parentheses, the standard deviation of a single run. True value $\chi=16$.}
\label{tab:three}
\end{table*}

Table~\ref{tab:three} shows the cost of weighting modes that the susceptibility does not depend on. The unweighted  estimate  agrees with  the prediction  $16/1.77=9.04$  of Proposition~\ref{prop:three}(ii) . Global weights give errors about 3 and 12 times larger than local weights for the relevant and irrelevant mismatches. With only $Ne^{-D_2}\approx20$ effective samples, a few large weights dominate and these ratios fluctuate strongly between sets of runs; Proposition~\ref{prop:three}(iii) gives asymptotic ratios $e^{(D_2-D_2(A))/2}=27$ and $30$.

A practical warning: $-\log(\widehat\ESS/N)$ can strongly underestimate $D_2$, because $\Eq[w^2]$ is dominated by rare large weights. For the relevant mismatch with $L=128$, $\xi=4$ and $\chat=0.50$, for which $D_2=11$, single runs with $N=3\times10^5$ give $-\log(\widehat\ESS/N)\approx7$ (median $6.95$ and $10$--$90\%$ range $6.5$--$8.2$ over 20 runs; see also \cite{Chatterjee:2018}). To compare samplers we recommend $\sigma^2/V=\mathrm{Var}_q(\log w)/V$, which uses typical samples and tests extensivity directly.

\subsection{Composite chain}\label{sec:exact}

In the free field the composite chain can be solved exactly. Take the Gaussian model of Sec.~\ref{sec:modes}, with $\Phi=x_0$ (the zero mode). One cycle is: (1) propose a whole configuration from $q$ and accept it with the marginal test on $x_0$; (2) run $\nref$ HMC trajectories with exact Hamiltonian flow. For each trajectory the momenta $\pi_k$ are drawn from $N(0,1)$ and one trajectory length $t$ is drawn from a distribution $P(t)$; the same $t$ is used for all modes, as in ordinary HMC. With $H=\sum_k(\pi_k^2+\lambda_kx_k^2)/2$, each mode rotates in phase space,
\begin{equation}\label{eq:rotation}
\begin{gathered}
x_k\ \to\ x_k\cos(\omega_kt)+\frac{\pi_k}{\omega_k}\sin(\omega_kt),\\
\omega_k=\sqrt{\lambda_k}=\sqrt{\kh^2+m^2}.
\end{gathered}
\end{equation}
The energy is conserved, so the Metropolis step at the end of each trajectory always accepts.

\begin{proposition}[Composite chain in the free field]\label{prop:exact}
In the stationary state:
\begin{enumerate}[label=(\roman*),leftmargin=2em]
\item The zero mode has exactly the target distribution, for any $\nref$. In particular $\chi$ has no bias.
\item For every $k\neq0$,
\begin{equation}\label{eq:survival}
\begin{gathered}
\frac{G_k^{\rm chain}}{G_k^{\rm true}}-1=\left(-\frac{\varepsilon_k}{1+\varepsilon_k}\right)\mathbb{E}[B_k],\\
B_k=\prod_{\substack{\text{trajectories since the}\\ \text{last accepted proposal}}}\cos^2(\omega_kt_i).
\end{gathered}
\end{equation}
The first factor is the unweighted bias of Proposition~\ref{prop:three}(ii); $B_k$ is the fraction of the model's error that has survived the local updates.
\item With $\bar a$ the mean acceptance rate and $\rho_k=\big(\mathbb{E}_t[\cos^2(\omega_kt)]\big)^{\nref}$,
\begin{equation}\label{eq:survival-bounds}
\bar a\,\rho_k\ \le\ \mathbb{E}[B_k]\ \le\ \rho_k.
\end{equation}
\item If $t$ is exponentially distributed with mean $\bar t$, then $\mathbb{E}_t[\cos^2(\omega t)]=(1+2\omega^2\bar t^2)/(1+4\omega^2\bar t^2)$, which decreases with $\omega$. Since $\omega_k\ge m$ for all $k$,
\begin{equation}\label{eq:taum}
\begin{gathered}
\max_{k\neq0}\left|\frac{G_k^{\rm chain}}{G_k^{\rm true}}-1\right|\ \le\ \max_k\left|\frac{\varepsilon_k}{1+\varepsilon_k}\right|\,e^{-\nref/\tau_m},\\
\tau_m=\left[\log\frac{1+4m^2\bar t^2}{1+2m^2\bar t^2}\right]^{-1}\simeq\frac{\xi^2}{2\bar t^2}\ \ (\bar t\ll\xi).
\end{gathered}
\end{equation}
At fixed $m$ and $\bar t$, this bound is independent of $L$ if the proposal's relative variance errors are bounded uniformly in volume.
\end{enumerate}
\end{proposition}

The proof is in Appendix~\ref{app:exact}. Each accepted proposal introduces the model's mode variances. An exact trajectory retains only the fraction $\cos^2(\omega_kt)$ of their error: high-frequency modes forget the proposal quickly, low-frequency modes slowly. The zero mode is handled by the marginal test, and all other frequencies satisfy $\omega_k\ge m$. Equation~\eqref{eq:taum} therefore bounds their relaxation by $\tau_m$, independently of volume. The bound is a maximum over modes, not a sum: although the number with $\kh\lesssim1/\xi$ grows as $(L/\xi)^d$, they relax in parallel. At fixed $\xi$ and a uniform bound on the relative errors in the proposal mode variances, $\nref$ need not grow with volume; the remaining slowing follows $\tau_m\simeq\xi^2/2\bar t^{\,2}$. Random trajectory lengths avoid resonances $\omega_kt=\pi,2\pi,\dots$ that would prevent relaxation~\cite{Mackenzie:1989}. For fixed $t$, replace $\mathbb{E}_t[\cos^2(\omega t)]$ by $\cos^2(\omega t)$: $\tau_m=[-\log\cos^2(mt)]^{-1}\simeq\xi^2/t^2$, provided $\omega_{\max}t<\pi/2$, with $\omega_{\max}=\sqrt{8+m^2}$.

If the acceptance decisions were independent from cycle to cycle, the time since the last acceptance would be geometric and $\mathbb{E}[B_k]=\bar a\rho_k/[1-(1-\bar a)\rho_k]$. In reality consecutive decisions are correlated through the zero mode, so this is an approximation, but it agrees with the exact result \eqref{eq:survival} to $0.5\%$ in the test below.

\begin{figure*}[t]
\centering
\includegraphics{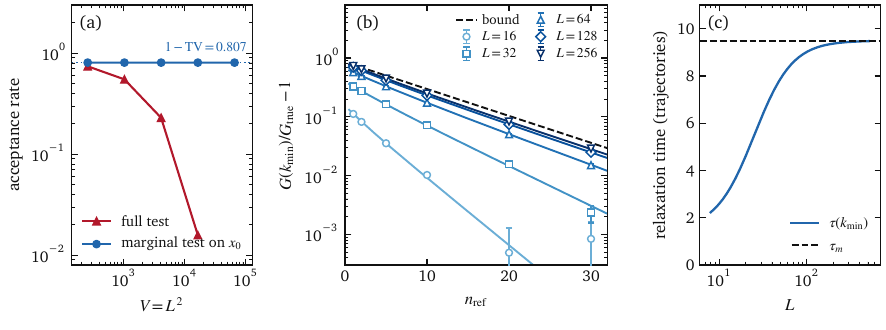}
\caption{Composite chain in the free field ($\xi=4$, $\varepsilon_0=-0.46$, $\bar t=1$, $10^4$ parallel chains). (a) The marginal test on $x_0$ accepts with the exact rate $0.807$ for all $L$ (points: measured; dotted line: $1-\TV$ from Proposition~\ref{prop:marg}); the full ratio test, shown with its exact acceptance rate, collapses. (b) Bias of $G(k_{\min})$ against $\nref$. Points: simulation; the two $L=16$ points at $\nref=20$ and $30$ are consistent with zero. Lines: Eq.~\eqref{eq:survival} with the geometric approximation for $\mathbb{E}[B_k]$. Dashed: the $L$ independent bound \eqref{eq:taum}. (c) Relaxation time $\tau(k_{\min})=-1/\log\mathbb{E}_t[\cos^2(\omega_{k_{\min}}t)]$ from Proposition~\ref{prop:exact}(iv): it grows with $L$ and saturates at $\tau_m$.}
\label{fig:free-composite}
\end{figure*}

\paragraph{Numerical check.} We simulated the chain for $L=16$ to $256$ with $\xi=4$ and $\varepsilon_k=\varepsilon_0m^2/(\kh^2+m^2)$, $\varepsilon_0=-0.46$. The model is then too close to criticality, and its unweighted $\chi$ bias is $+85\%$. Since the modes are independent, it is enough to follow $x_0$ and the four real modes at $k_{\min}$; this is exact, not an approximation. Figure~\ref{fig:free-composite} shows the result.
(a) The marginal test accepts at $0.807$ for every $L$, equal to the exact value $0.8069$ from Proposition~\ref{prop:marg}; the full ratio test gives $0.742$, $0.551$, $0.229$, $0.016$ for $L=16$, $32$, $64$, $128$.
(b) The zero mode variance agrees with the target within errors in all 30 runs (largest deviation $2.1\sigma$), as stated in (i). The measured $G(k_{\min})$ agrees with Eq.~\eqref{eq:survival}, evaluated with the $\mathbb{E}[B_k]$ measured in the same run, with $\chi^2/{\rm dof}=0.90$ over 30 points.
(c) At fixed $\nref$ the bias grows with $L$ and then saturates. The reason is visible in Eq.~\eqref{eq:survival}: as $L$ grows, $k_{\min}$ approaches $0$, so both the unweighted bias ($0.15$, $0.40$, $0.66$, $0.80$, $0.84$) and the relaxation time ($3.8$, $6.4$, $8.4$, $9.2$, $9.4$ trajectories) increase, but both stop at the values of a mode with frequency $m$, $0.85$ and $\tau_m=9.5$. For a target accuracy $\delta$ one needs $\nref\ge\tau_m\log(0.85/\delta)$, about 42 trajectories for $\delta=1\%$, for any $L$.

\paragraph{Scope of the free-field analysis.} The bound concerns the residual mode-variance bias at fixed $\xi$. Three effects remain to be tested. A leapfrog integrator couples mode acceptance through the energy error, unlike exact Hamiltonian flow. For $L\gg\xi$, the zero mode is not much slower than the others, so this example does not test the benefit of moving it separately. Finally, interactions couple the zero mode to the other modes and can disturb its distribution during relaxation. Section~\ref{sec:phi4chain} examines these effects near criticality in $\phi^4$ theory: proposals with excess short-distance fluctuations pull the zero mode during relaxation, an effect suppressed by relaxing the proposal before the test.

\section{Two-dimensional \texorpdfstring{$\phi^4$}{φ⁴} theory}\label{sec:phi4}

\begin{figure*}[t]
\centering
\includegraphics{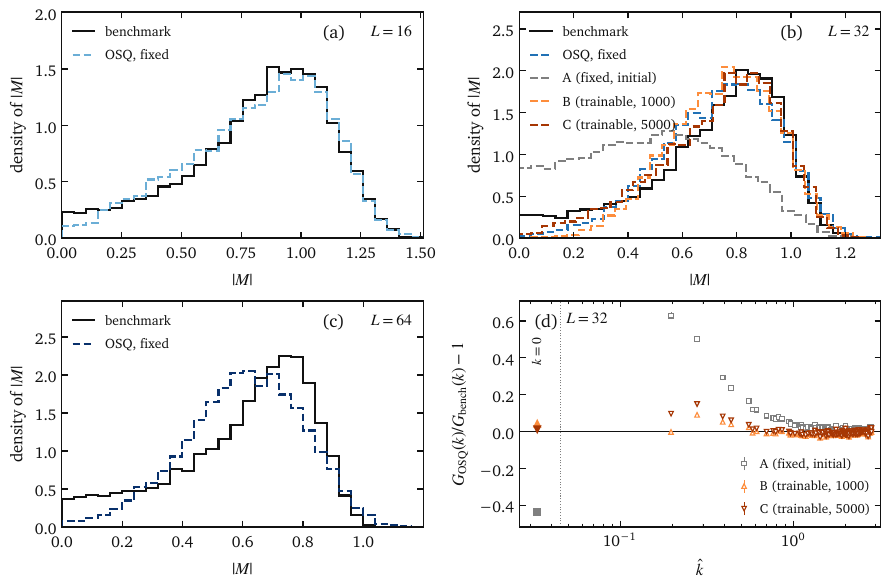}
\caption{OSQ samplers at $\kappa=0.2705$ compared with the benchmark ensembles. (a)--(c) Distribution of $|M|$ from the benchmark ensembles (solid) and from unweighted OSQ samples (dashed) with fixed reference (Table~\ref{tab:osq-samplers}) at $L=16$, $32$ and $64$; panel (b) also shows the samplers A--C of Table~\ref{tab:osq-diag}. (d) Relative deviation $G_{\rm OSQ}(k)/G_{\rm bench}(k)-1$ of the OSQ propagator from the benchmark against the lattice momentum $\kh$ for the samplers A--C, averaged over lattice-symmetry orbits; the filled symbols at the left are the zero mode, $G(0)=\chi_0$.}
\label{fig:osq}
\end{figure*}

We test the picture of Secs.~\ref{sec:ess}--\ref{sec:free} near criticality with trained OSQ samplers~\cite{Wang:OSQ} at $L=16$, $32$ and $64$, with independent high-precision Monte Carlo ensembles as benchmark. Figures~\ref{fig:osq} and~\ref{fig:osq-volume} compare magnetization and Fourier-mode errors with global overlap. Varying two reference-process parameters tests whether the infrared observables can change substantially while the log-weight variance stays nearly fixed. We then use both trained samplers and their reference processes as proposals  in the composite chain, with the target distribution of the collective variables supplied as input,  and measure how  far  local updates  reduce  the remaining errors (Sec.~\ref{sec:phi4chain}). As in Sec.~\ref{sec:free}, Secs.~\ref{sec:hmc-osq} and~\ref{sec:phi4-ir} concern the first question of the introduction and Sec.~\ref{sec:phi4chain} the second.

\subsection{Lattice setup and samplers}\label{sec:phi4-setup}
The lattice action is
\begin{equation}\label{eq:phi4action}
S=\sum_x\Big[-2\kappa\sum_\mu\phi_x\phi_{x+\hat\mu}+\phi_x^2+\lambda\big(\phi_x^2-1\big)^2\Big],
\end{equation}
with $\lambda=0.022$ and $\kappa=0.2705$, close to the critical coupling $\kappa_c=0.27088$~\cite{Tan:2026}. As benchmark we use the ensembles of Ref.~\cite{Tan:2026}, which are publicly available~\cite{Tan:2026data}. They consist of 10240 configurations per volume, generated with Fourier-accelerated HMC combined with Wolff cluster updates, and the integrated autocorrelation times of the observables used here are below $0.8$ configurations~\cite{Tan:2026data}. We call them the benchmark ensembles, to distinguish them from the plain (not Fourier-accelerated) HMC that serves as the exact local update in Sec.~\ref{sec:phi4chain}. Since two-dimensional $\phi^4$ theory is not exactly solvable, all accuracies quoted below are relative deviations $X/X_{\rm bench}-1$ from Monte Carlo benchmark values; where errors are given, they include those of the benchmark, and agreement within errors means consistency with the benchmark rather than exactness. We measure the magnetization $M=V^{-1}\sum_x\phi_x$, the propagator $G(k)=\langle|\tilde\phi(k)|^2\rangle/V$ with $\tilde\phi(k)=\sum_xe^{-ik\cdot x}\phi_x$, the susceptibility $\chi_0=V\langle M^2\rangle=G(0)$, which is the quantity called susceptibility in the free-field analysis, the connected susceptibility $\chi=V(\langle M^2\rangle-\langle|M|\rangle^2)$ used in Ref.~\cite{Tan:2026}, the Binder cumulant $U_L=1-\langle M^4\rangle/(3\langle M^2\rangle^2)$, $\langle\phi_x^2\rangle$ and $\langle S\rangle/V$. Errors are from a jackknife analysis with 64 blocks. At this coupling the correlation length is limited by the box: the second-moment correlation length $\xi=[\chi_0/G(k_{\min})-1]^{1/2}/[2\sin(\pi/L)]$, with $k_{\min}=2\pi/L$, is $0.76L$, $0.75L$ and $0.66L$ at $L=16$, $32$ and $64$.

An OSQ sampler is a diffusion sampler whose drift is optimized by stochastic optimal control. It integrates the stochastic differential equation
\begin{equation}\label{eq:osq-sde}
d\phi_t=\big[u_{\rm ref}(\phi_t)+u_\theta(\phi_t,t)\big]\,dt+g\,dW_t,\qquad 0\le t\le T,
\end{equation}
starting from the stationary distribution $p_{\rm ref}$ of a reference process with drift $u_{\rm ref}$. The reference process is a free Ornstein--Uhlenbeck process, $u_{\rm ref}(\phi)=-K_{\rm ref}(\phi-c)$, where $K_{\rm ref}$ has the Fourier kernel $\hat A(k)+m_{\rm ref}$ and $\hat A(k)=2(1-2\lambda)-4\kappa\sum_\mu\cos k_\mu$ is the kernel of the quadratic part of Eq.~\eqref{eq:phi4action}. At our couplings $\hat A(0)=-0.252$, and the mass shift $m_{\rm ref}>0.252$ makes the process stable. The constant field $c$ equals $+\mu_{\rm ref}$ or $-\mu_{\rm ref}$ with probability $1/2$. In $p_{\rm ref}$ the nonzero modes are therefore Gaussian with variance close to $1/[\hat A(k)+m_{\rm ref}]$ (for $g^2=2$), and $M$ has the double-peaked distribution $\frac12\mathcal N(\mu_{\rm ref},\sigma_0^2)+\frac12\mathcal N(-\mu_{\rm ref},\sigma_0^2)$ with $\sigma_0^2\simeq1/[V(\hat A(0)+m_{\rm ref})]$. The residual drift $u_\theta$ is a convolutional network (a U-Net with periodic padding and a Fourier embedding of $t$) that is odd under $\phi\to-\phi$. It is trained by minimizing
\begin{equation}\label{eq:osq-loss}
\mathcal L(\theta)=\mathbb E\Big[\int_0^T\frac{|u_\theta|^2}{2g^2}\,dt+S(\phi_T)+\log p_{\rm ref}(\phi_T)\Big],
\end{equation}
which equals $D_{\rm KL}(Q_\theta\|P)-\log Z$~\cite{Wang:OSQ}. Here $Q_\theta$ is the path measure of Eq.~\eqref{eq:osq-sde}, and $P$ is the path measure of the reference process reweighted by $e^{-S(\phi_T)}/[Z\,p_{\rm ref}(\phi_T)]$, whose distribution at $t=T$ is the target $p$. Each sample comes with the path weight
\begin{equation}\label{eq:pathweight}
\begin{split}
\log w={}&-S(\phi_T)-\log p_{\rm ref}(\phi_T)\\
&-\int_0^T\frac{|u_\theta|^2}{2g^2}\,dt-\int_0^T\frac{u_\theta\cdot dW_t}{g}.
\end{split}
\end{equation}
The conditional mean of $w$ over paths ending at $\phi_T$ is proportional to $p(\phi_T)/q(\phi_T)$, where $q$ is the final-field distribution. The OSQ ESS and full-test rates below use these path weights, which can fluctuate even when the final fields are accurate. We use $T=10$, $g=\sqrt2$ and 250 Euler--Maruyama steps, as in training, and $p_{\rm ref}$ is the stationary distribution of the discretized process.

The infrared of an OSQ sampler is controlled mainly by the two parameters $\mu_{\rm ref}$ and $m_{\rm ref}$ of the reference process, which fix the distribution of the zero mode in $p_{\rm ref}$. We first fix them from the benchmark values, $\mu_{\rm ref}=\langle|M|\rangle$ and $m_{\rm ref}=-\hat A(0)+1/\chi$, rounded, and train one sampler per volume with the same network and protocol and 1000 training steps (Table~\ref{tab:osq-samplers}); samplers in which the two parameters are trained follow in Sec.~\ref{sec:phi4-ir}. Since $\sigma_0^2\simeq1/[V(\hat A(0)+m_{\rm ref})]$, this gives $V\sigma_0^2\simeq\chi$ and $V\langle M^2\rangle_{\rm ref}\simeq\chi_0$: the reference carries the statistics of the zero mode, and the network has to supply the rest. Table~\ref{tab:osq-samplers} compares these quantities in $p_{\rm ref}$ with the benchmark values; after the rounding, $\langle|M|\rangle$ and $\chi_0$ agree within $5\%$ and $\chi$ within $15\%$. From each of these samplers, and from those of Sec.~\ref{sec:phi4-ir}, we draw $N=10240$ independent configurations.

\begin{table*}[t]
\centering\small
\begin{tabular}{ccccccccc}
\toprule
 & & & \multicolumn{3}{c}{$p_{\rm ref}$} & \multicolumn{3}{c}{benchmark}\\
\cmidrule(lr){4-6}\cmidrule(lr){7-9}
$L$ & $m_{\rm ref}$ & $\mu_{\rm ref}$ & $\langle|M|\rangle$ & $\chi$ & $\chi_0$ & $\langle|M|\rangle$ & $\chi$ & $\chi_0$\\
\midrule
16 & 0.30 & 0.80 & 0.80 & 21 & 185 & 0.8149(26) & 22.99(30) & 192.98(94)\\
32 & 0.27 & 0.72 & 0.72 & 55 & 586 & 0.7147(25) & 64.28(83) & 587.3(32)\\
64 & 0.257 & 0.60 & 0.60 & 198 & 1675 & 0.6019(26) & 211.8(30) & 1696(11)\\
\bottomrule
\end{tabular}
\caption{OSQ samplers with fixed reference at $\kappa=0.2705$. All have the same network with $6.7\times10^4$ parameters and use $T=10$, $g=\sqrt2$, 250 Euler--Maruyama steps, batches of 16 paths, 1000 training steps with a cosine learning-rate schedule and an exponential moving average of the parameters. The reference parameters $m_{\rm ref}$ and $\mu_{\rm ref}$ are fixed from the benchmark values, rounded; the training objective~\eqref{eq:osq-loss} uses no benchmark configurations. The columns $p_{\rm ref}$ give $\langle|M|\rangle$, $\chi$ and $\chi_0$ in the reference distribution, the columns benchmark those of the benchmark ensembles.}
\label{tab:osq-samplers}
\end{table*}

\subsection{Infrared accuracy and global overlap}\label{sec:hmc-osq}

\begin{table*}[t]
\centering\small
\setlength{\tabcolsep}{6pt}
\begin{tabular}{clllllll}
\toprule
$L$ & & $\langle|M|\rangle$ & $\chi_0$ & $\chi$ & $U_L$ & $\langle\phi_x^2\rangle$ & $\langle S\rangle/V-\lambda$\\
\midrule
16 & benchmark & 0.8149(26) & 192.98(94) & 22.99(30) & 0.5524(18) & 1.5215(29) & 0.38758(49) \\
   & OSQ & 0.8150(30) & 191.9(11) & 21.84(33) & 0.5492(20) & 1.5378(31) & 0.38589(47) \\
   & OSQ, rw & 0.8164(64) & 192.8(22) & 22.22(74) & 0.5578(39) & 1.5205(69) & 0.3877(11) \\
\midrule
32 & benchmark & 0.7147(25) & 587.3(32) & 64.28(83) & 0.5672(16) & 1.4698(19) & 0.39297(26) \\
   & OSQ & 0.7351(16) & 600.9(23) & 47.55(72) & 0.5777(13) & 1.4905(16) & 0.38909(19) \\
   & OSQ, rw & 0.720(18) & 589(26) & 58.0(53) & 0.569(10) & 1.476(13) & 0.3936(30) \\
\midrule
64 & benchmark & 0.6019(26) & 1696(11) & 211.8(30) & 0.5551(20) & 1.4400(11) & 0.39619(18) \\
   & OSQ & 0.6024(17) & 1640.8(81) & 154.6(17) & 0.5587(13) & 1.4440(10) & 0.39365(15) \\
   & OSQ, rw$^\dagger$ & 0.610(54) & 1640(230) & 121(47) & 0.601(50) & 1.426(11) & 0.3986(16) \\
\bottomrule
\end{tabular}
\caption{Observables at $\kappa=0.2705$ from the benchmark ensembles and from the OSQ samplers of Table~\ref{tab:osq-samplers}, without weights and reweighted with the path weights~\eqref{eq:pathweight} (rw). $\chi_0=V\langle M^2\rangle$, $\chi=V(\langle M^2\rangle-\langle|M|\rangle^2)$. Errors from a jackknife analysis with 64 blocks. $^\dagger$The effective sample size is about 6 and the largest weight carries a third of the total weight, so these errors are not reliable.  At $L=32$ the weights are also too heavy-tailed for reliable rw errors (Appendix~\ref{app:pphi}). }
\label{tab:osq-obs}
\end{table*}

\begin{figure}[t]
\centering
\includegraphics[width=\columnwidth]{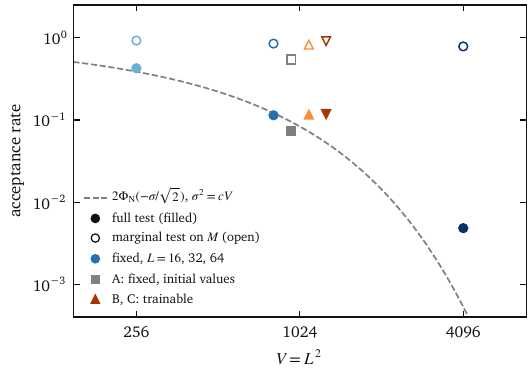}
\caption{Acceptance rates against the volume. Filled symbols: independence test with the path weights, Eq.~\eqref{eq:alpha-TV}; open symbols: marginal test on $M$, Eq.~\eqref{eq:margtest}; both are estimated from the samples as in Table~\ref{tab:osq-overlap}, with $p_M$ taken from the benchmark ensembles. Circles: fixed reference (Table~\ref{tab:osq-samplers}) at $L=16$, $32$ and $64$; square: A; triangles: trainable reference (B, C). The samplers at $L=32$ are drawn slightly displaced in $V$. The dashed line is the Gaussian value $2\Phi_{\rm N}(-\sigma/\sqrt2)$ with $\sigma^2=cV$ and $c=5.9\times10^{-3}$, the mean of $\sigma^2/V$ of the samplers of Table~\ref{tab:osq-samplers}. The full-test value at $L=64$ is dominated by a few samples and may overestimate the true rate.}
\label{fig:osq-volume}
\end{figure}

Figure~\ref{fig:osq}(a)--(c) locates the main error: the sampler produces too few configurations with small $|M|$, where large regions of both signs coexist. At $L=64$, $|M|<0.2$ occurs in $2.7\%$ of OSQ samples against $8.1\%$ of the benchmark configurations. Consequently $\chi=V\,\mathrm{Var}(|M|)$ lies about a quarter below the benchmark at $L=32$ and $64$ ($5\%$ at $L=16$), even though $\langle|M|\rangle$, $\chi_0$ and $U_L$  lie  within about $3\%$  of their benchmark values  (Table~\ref{tab:osq-obs}; the first two are close to the values built into $p_{\rm ref}$, Table~\ref{tab:osq-samplers}). The local quantities $\langle\phi_x^2\rangle$ and $\langle S\rangle/V$  deviate by at most  $1.5\%$ , although these small deviations exceed the statistical errors. For $\kh>1$ the root-mean-square relative deviation of the propagator from the benchmark is about $1\%$, comparable to the statistical errors, while the lowest nonzero modes deviate by at most $15\%$. Accurate low moments therefore do not imply an accurate magnetization distribution.  Reweighting with the path weights removes this discrepancy at $L=16$ but not reliably at $L=32$ and $64$, where the weights are too heavy-tailed (Appendix~\ref{app:pphi}). 

\begin{table}[t]
\centering\small
\setlength{\tabcolsep}{1.8pt}
\begin{tabular}{cccccccc}
\toprule
$L$ & $\sigma^2$ & $10^3\,\sigma^2/V$ & $\ESS/N$ & $\alpha$ & $\alpha_M$ & $D_2(p_M\|q_M)$ & $D_2^{k\neq0}$\\
\midrule
16 & 1.38 & 5.4 & 0.30 & 0.43 & 0.92 & 0.03 & 0.05\\
32 & 5.34 & 5.2 & 0.019 & 0.11 & 0.85 & 0.18 & 0.07\\
64 & 28.5 & 7.0 & $6\times10^{-4}{}^\dagger$ & $0.005^\dagger$ & 0.78 & 0.21 & 0.46\\
\bottomrule
\end{tabular}
\caption{Global and infrared overlap of the samplers with fixed reference (Table~\ref{tab:osq-samplers}). $\sigma^2=\mathrm{Var}(\log w)$, $\ESS/N$ [Eq.~\eqref{eq:ESS-D2}] and the acceptance rate $\alpha$ of the independence test [Eq.~\eqref{eq:alpha-TV}, estimated from all pairs of samples] are computed with the path weights~\eqref{eq:pathweight}. $\alpha_M$ is the acceptance rate of the marginal test~\eqref{eq:margtest} with $\Phi=M$, and $D_2(p_M\|q_M)$ is the R\'enyi divergence of the distributions of $M$; both are obtained from histograms of $|M|$ (40 bins of equal total occupation) of the benchmark and the unweighted OSQ samples. $D_2^{k\neq0}$ is the Gaussian proxy for the nonzero modes defined in the text. $^\dagger$Dominated by a few samples; the Gaussian values are $e^{-\sigma^2}=4\times10^{-13}$ and $2\Phi_{\rm N}(-\sigma/\sqrt2)=2\times10^{-4}$.}
\label{tab:osq-overlap}
\end{table}

The global overlap, in contrast, collapses with the volume (Table~\ref{tab:osq-overlap} and Fig.~\ref{fig:osq-volume}). The variance of the log weights grows approximately in proportion to $V$ over these volumes, with $\sigma^2/V$ between $5$ and $7\times10^{-3}$ at all three volumes, so that $\ESS/N$ falls from $0.30$ at $L=16$ to $6\times10^{-4}$ at $L=64$ and the acceptance rate $\alpha$ of the independence test from $0.43$ to $0.005$. Where $\alpha$ can be measured reliably, at $L=16$ and $32$, it is close to the Gaussian estimate $2\Phi_{\rm N}(-\sigma/\sqrt2)=0.41$ and $0.10$ of Sec.~\ref{sec:IR}; at $L=64$ both estimates are dominated by a few samples and, as in the free field (Sec.~\ref{sec:estimators}), may overestimate the true values if the finite sample misses rare large weights. A variance proportional to $V$ is what contributions spread over the whole lattice produce: $\log w$ is then a sum of many weakly correlated terms, as for the irrelevant mismatch of the free field (Sec.~\ref{sec:free}). We  examine their origin  below.  The falling effective sample size limits reweighting with the path weights: at $L=16$ and $32$ it brings all observables of Table~\ref{tab:osq-obs} within about two standard deviations of the benchmark, at the cost of errors larger by factors of about 2 and 7 to 15,  but at $L=32$ these errors are themselves unreliable (Appendix~\ref{app:pphi}), and  at $L=64$, with an effective sample size of about 6, the reweighted $\chi$ is far from the benchmark value.

The infrared overlap, on the other hand, stays of order one. Between $L=16$ and $64$ the acceptance rate of the marginal test on $M$ decreases only from $0.92$ to $0.78$, reflecting the larger infrared error of these samplers at larger $L$, while that of the global test falls by two orders of magnitude (Fig.~\ref{fig:osq-volume}), and $\alpha_M\ge\alpha$ in all cases, as required by Proposition~\ref{prop:marg}. The R\'enyi divergence of the distributions of $M$ and the Gaussian proxy $D_2^{k\neq0}=-\frac12\sum_{k\neq0}\log(1-\varepsilon_k^2)$ of the divergence of the nonzero modes, with $1+\varepsilon_k=G_{\rm bench}(k)/G_{\rm OSQ}(k)$ [cf.\ Eq.~\eqref{eq:D2free}] after subtraction of the statistical bias, are both below $0.5$ at all three volumes (Table~\ref{tab:osq-overlap}), far below $\sigma^2$. These diagnostics suggest that the resolved infrared mismatch contributes little to the variance of the log weights. Here $p_M$ is taken from the benchmark ensemble and serves only as a diagnostic; in a sampler it has to be determined independently (Sec.~\ref{sec:fix}).

At $L=64$, the distinction is particularly clear: the sampler reproduces $\langle|M|\rangle$, $\chi_0$ and $U_L$ to $3\%$, yet has about 6 effective samples out of $10240$. A test on $M$ would accept about $0.8$ of its proposals and correct that marginal up to density-estimation errors (Sec.~\ref{sec:phi4chain}).

For OSQ samplers, much of $\sigma^2$ appears to come from the generating paths. Besides the zero-mode divergence and nonzero-mode estimate, shifts in the local operators $\phi_x^2$, $\phi_x^4$ and $\sum_\mu\phi_x\phi_{x+\hat\mu}$ give $D_2\lesssim0.2$ at $L=32$ and $64$ in the approximation~\eqref{eq:gibbs}, far below $\sigma^2=5.3$ and $28.5$. These checks can miss higher field moments~\cite{Aarts:2024b} and correlations, so they do not establish a complete decomposition. Path weights describe fields in $d+1$ dimensions, including fictitious time, and can fluctuate even for an exact endpoint distribution unless the drift is optimal. Projecting a path onto $\phi_T$ is a coarse graining: by Sec.~\ref{sec:flow}, $D_2(p\|q)$ cannot exceed the path divergence. The population ESS fraction from path weights therefore bounds the endpoint ESS fraction from below; here the bound appears far from saturated.

\subsection{Fixed and trainable reference parameters}\label{sec:phi4-ir}
If the log-weight variance is dominated by contributions outside the infrared, changing mainly the infrared of a sampler can change its long-distance observables while leaving this variance almost unchanged. We test this at $L=32$ by training the two reference parameters together with the network, which also avoids fixing them from the target with information that a sampler does not have in general. The objective~\eqref{eq:osq-loss} can be used unchanged, since it equals $D_{\rm KL}(Q_\theta\|P)-\log Z$ for every reference process. The two parameters are initialized by the rule of Ref.~\cite{Wang:OSQ}, which is based on the mean-field vacuum $\sqrt{-\hat A(0)/4\lambda}$ and gives $m_{\rm ref}=0.264$ and $\mu_{\rm ref}=0.40$ at $L=32$. The gradient reaches them through the initial sample $\phi_0\sim p_{\rm ref}$, through the reference drift and through $\log p_{\rm ref}(\phi_T)$, which must now include the normalization of every mode. The parametrization $m_{\rm ref}=-\hat A(0)+e^{\rho}$ keeps the reference stable, and the two parameters use a learning rate five times that of the network. As for the other samplers, all gradients are evaluated with the reference drift held fixed along the path. Table~\ref{tab:osq-diag} compares three samplers with the network of Table~\ref{tab:osq-samplers}: A, with the reference fixed at these initial values and 1000 training steps, and B and C, one run with trainable reference after 1000 and 5000 training steps. The learning rates follow a cosine schedule over the 5000 steps, so B is taken at $90\%$ of the initial learning rate, whereas A has completed its schedule.

\begin{table*}[t]
\centering\small
\setlength{\tabcolsep}{3.5pt}
\begin{tabular}{lccccccccccc}
\toprule
 & steps & $m_{\rm ref}$ & $\mu_{\rm ref}$ & $\langle|M|\rangle$ & $\chi_0$ & $\chi$ & $10^3\,\sigma^2/V$ & $\ESS/N$ & $\alpha$ & $\alpha_M$ & $D_2(p_M\|q_M)$\\
\midrule
\multicolumn{4}{l}{benchmark} & 0.7147(25) & 587.3(32) & 64.28(83) & & & & & \\
A & 1000 & 0.264 & 0.403 & 0.4977(21) & 330.4(25) & 76.75(96) & 5.01 & 0.003 & 0.073 & 0.545 & 0.613 \\
B & 1000 & 0.283 & 0.750 & 0.7493(14) & 612.7(21) & 37.82(56) & 5.02 & 0.023 & 0.116 & 0.824 & 0.715 \\
C & 5000 & 0.286 & 0.725 & 0.7274(16) & 594.0(22) & 52.10(79) & 5.00 & 0.027 & 0.118 & 0.907 & 0.063 \\
\bottomrule
\end{tabular}
\caption{Samplers at $L=32$ with the reference held fixed at the initial values $m_{\rm ref}=0.264$, $\mu_{\rm ref}=0.403$ (A) and with trainable reference (B, C), after the given number of training steps; for B and C, $m_{\rm ref}$ and $\mu_{\rm ref}$ are the learned values. Observables without weights; the other quantities are defined as in Table~\ref{tab:osq-overlap}.}
\label{tab:osq-diag}
\end{table*}

The zero modes of the three samplers differ qualitatively [Fig.~\ref{fig:osq}(b)], but the variance of their log weights is the same within $1\%$, $\sigma^2/V=5.0\times10^{-3}$. For A, training does not repair the zero mode: its distribution of $|M|$ is shifted to small values, so that $\langle|M|\rangle$ lies $30\%$ below the benchmark and $\chi_0$ is little more than half the benchmark value, and the four lowest nonzero modes exceed the benchmark by $24$--$63\%$ [Fig.~\ref{fig:osq}(d)]. The global diagnostics respond to this error but do not locate it: $\alpha$ and $\ESS/N$ of A are smaller than those of B and C, but both estimates depend on the tail of the weight distribution (for A the effective sample size is about 30, and the largest weight carries a sixth of the total), and neither says where the mismatch is. The marginal test does: $\alpha_M=0.55$ for A, against $0.82$ and $0.91$ for B and C, and $D_2(p_M\|q_M)=0.61$ comes mainly from $|M|>0.9$, where $q_M$ is too small. Even for A, a marginal test would thus correct the zero mode with an acceptance rate of order one.

In the run with trainable reference parameters,  the typical magnitude of $M$ is learned, although not monotonically: $\mu_{\rm ref}$ reaches the benchmark value $\langle|M|\rangle=0.715$ within about 400 training steps, overshoots and fluctuates while the learning rate is large, and settles at $0.72$. The parameter $\mu_{\rm ref}$ enters only the zero mode of the reference, so that its gradient is not drowned by the ultraviolet contributions that dominate the objective; for the parameters of the network, which act on all modes, the infrared is only a small part of the gradient, and this is presumably why training does not repair the zero mode of A. B, taken after 1000 steps with $\mu_{\rm ref}=0.75$, has $\langle|M|\rangle$ and $\chi_0$ only a few percent above the benchmark, but its distribution of $|M|$ is too narrow [Fig.~\ref{fig:osq}(b)]: configurations with small $|M|$ are almost absent ($|M|<0.2$ in $0.3\%$ of its samples against $5.3\%$ for the benchmark), and $\chi$ is $41\%$ below the benchmark. The two infrared diagnostics weigh this  discrepancy  differently: $\alpha_M=0.82$ shows that the bulk of the distribution overlaps well, whereas $D_2(p_M\|q_M)=0.72$ is even larger than for A, because $D_2$ is dominated by the regions where $q_M\ll p_M$, here those near $M=0$. After 5000 steps (C), $\langle|M|\rangle$ and $\chi_0$  lie  within $2\%$  of the benchmark, and $D_2(p_M\|q_M)=0.06$ is smaller than for the sampler with fixed reference at $L=32$ (Table~\ref{tab:osq-overlap}); as for that sampler, the remaining error is the width of the distribution of $|M|$, with $\chi$ $19\%$ below the benchmark.

The comparisons show why the magnetization distribution and low-momentum propagator are needed alongside the ESS. The marginal acceptance stays high as the global ESS falls; changing the reference parameters alters infrared observables while leaving the path log-weight variance nearly unchanged. These infrared diagnostics require only unweighted model samples and reference information on the chosen physical quantities.

\subsection{Ultraviolet correction by exact local updates}\label{sec:phi4chain}
Sections~\ref{sec:hmc-osq} and~\ref{sec:phi4-ir} indicate that the trained samplers' ESS is suppressed mainly by contributions outside the infrared: path-weight fluctuations and a small mismatch spread over many short-distance modes, with little effect on long-distance observables.  Given the target magnetization marginal, we now examine the errors that remain in full configurations and the cost of relaxing them.  Global reweighting has a nominal effective-sample cost of order $N/\ESS$, while an independence test needs $1/\alpha$ proposals per acceptance; both can grow exponentially with $V$ (Sec.~\ref{sec:hmc-osq}). The composite chain of Sec.~\ref{sec:fix} divides the task: the marginal test alone targets the exact zero-mode distribution when the marginal densities are known, and exact local updates, plain HMC trajectories at the target, reduce the remaining mismatch.  The cost per effective independent measurement depends on both the work per cycle and the autocorrelation time of the observable; Table~\ref{tab:cost} uses $M^2$ to measure decorrelation of the zero mode.  In the free field at fixed $\xi$ the number of trajectories needed does not grow with $V$ (Sec.~\ref{sec:exact}). Here we measure the cost in $\phi^4$ theory at $\kappa=0.2705$, where we examine two effects absent from the free-field test of Sec.~\ref{sec:exact}, at fixed $\xi=4$: interactions couple the zero mode to the short-distance fluctuations, and the lowest nonzero modes relax slowly because $\xi$ is comparable to $L$, as expected for the local dynamics used here (Sec.~\ref{sec:fix}). Since the correlation length at this coupling is limited by the box (Sec.~\ref{sec:phi4-setup}), increasing $L$ increases $\xi$ as well, so this comparison alone does not separate dependence on $\xi$ from dependence on $V$. The volume dependence at fixed $\xi$, the regime of Proposition~\ref{prop:exact}, is studied in Appendix~\ref{app:fixedxi} at $\kappa=0.264$, with $\xi\simeq5$ and $L$ up to $256$.

\paragraph{Proposals and acceptance.} We use two kinds of proposals. The first is the stationary distribution of the continuous-time OSQ reference process, without the network or time discretization,
\begin{equation}\label{eq:qref}
\begin{gathered}
q(\phi)=\tfrac12\,\mathcal N\big(\phi;+\mu_{\rm ref},K_{\rm ref}^{-1}\big)+\tfrac12\,\mathcal N\big(\phi;-\mu_{\rm ref},K_{\rm ref}^{-1}\big),\\
K_{\rm ref}(k)=\hat A(k)+m_{\rm ref},
\end{gathered}
\end{equation}
sampled exactly by a fast Fourier transform, with $m_{\rm ref}$ and $\mu_{\rm ref}$ from Table~\ref{tab:osq-samplers} and, at $L=32$, also with the initial values of sampler A of Table~\ref{tab:osq-diag}. Both $\log q$ and $q_M$ are known in closed form, allowing the full test to be computed exactly and compared with the marginal test. The infrared of this proposal is set by the same two parameters as in the trained samplers, while its ultraviolet is poor: $\langle S\rangle/V$ is too large by $8$--$14\%$. It exposes a large ultraviolet error that the network removes in trained samplers. Proposals of the second kind come from the trained samplers of Tables~\ref{tab:osq-samplers} and~\ref{tab:osq-diag} themselves, whose ultraviolet is already accurate (Sec.~\ref{sec:hmc-osq}). Each proposal is used only once within a run, drawn from 20480 fresh samples per sampler. For these samplers $q_M$ is obtained by kernel density estimation from 2000--4000 proposals.

One cycle follows Sec.~\ref{sec:fix}: propose $\phi'\sim q$, accept it with the marginal test~\eqref{eq:margtest} on $\Phi=M$, then run $\nref$ trajectories of plain (not Fourier-accelerated) HMC at the target, with leapfrog step $\epsilon=0.1$, trajectory length $t\sim{\rm Exp}(1)$ and a Metropolis step at the end. For $p_M$ we use a symmetrized kernel density estimate from our own ensembles, generated with plain HMC and Wolff cluster updates in the embedding of Ref.~\cite{Brower:1989}.  The target marginal is thus an input: observables of $M$ show how well the marginal test and the local updates reproduce it, whereas $G(k_{\min})$ and $\langle S\rangle/V$ probe the modes outside $\Phi$.  Each point uses 16 independent chains of 200--1500 cycles (220 for the trained samplers) after 30 cycles of equilibration, and the errors include those of the benchmark values. Two independent implementations agree within statistical errors.\footnote{Each implementation has its own ensembles generated with plain HMC and Wolff updates, 8 chains of $2$--$2.4\times10^4$ configurations in the first and of 3000--6000 configurations in the second; at $L=32$ and $64$ their observables agree with the benchmark values of Table~\ref{tab:osq-samplers} within errors. For the 74 deviations of $\chi_0$ and $G(k_{\min})$ computed with both implementations and the same proposals, the difference is on average $0.85$ standard deviations and exceeds two in 4 cases. The results for the reference process come from the first implementation, with benchmark values from its own ensembles, and those for the trained samplers from the second; for the latter, the deviations of $\chi_0$, $\chi$ and $\langle|M|\rangle$ are taken with respect to the more precise benchmark values of Table~\ref{tab:osq-samplers}, and for the sampler of Table~\ref{tab:osq-samplers} at $L=64$ two sets of 16 chains are combined.}

The first cost is the number of proposals per acceptance (Table~\ref{tab:chainacc}). For the reference process the full test, which needs the density ratio of the whole configuration, accepts $0.13$ of the proposals at $L=16$ and less than $10^{-10}$ at $L=64$, whereas the marginal test accepts about $0.8$ of them at all three volumes. The reference process has nearly the right zero mode but a poor ultraviolet, and its log weights collect this mismatch from the whole lattice, making $\sigma^2/V$ much larger than for trained samplers; it grows with $L$ , presumably  because, with the rounded parameters of Table~\ref{tab:osq-samplers}, the error of the low modes changes with $L$ near criticality. The trained samplers are accepted by the marginal test more often than their reference processes, at rates close to the histogram estimates of $\alpha_M$ in Sec.~\ref{sec:hmc-osq}. With the marginal test they thus need fewer than $1.25$ proposals per accepted one at every volume, whereas with the full test, using the path weights, they would need from about 2 at $L=16$ to more than 200 at $L=64$ (Table~\ref{tab:osq-overlap}).

\begin{table*}[t]
\centering\small
\setlength{\tabcolsep}{4pt}
\begin{tabular}{lccccc}
\toprule
proposal & $L$ & $\sigma^2/V$ & full test $\alpha$ & marginal test $\alpha_M$ & trained sampler $\alpha_M$\\
\midrule
reference of Table~\ref{tab:osq-samplers} & 16 & 0.030 & 0.13 & 0.79 & 0.93\\
                     & 32 & 0.13 & $1.2\times10^{-3}$ & 0.78 & 0.86--0.88\\
                     & 64 & 0.85 & $<10^{-10}$ & 0.76 & 0.82--0.83\\
reference of A (Table~\ref{tab:osq-diag}) & 32 & 0.038 & $1.6\times10^{-3}$ & 0.40 & 0.54--0.56\\
\bottomrule
\end{tabular}
\caption{Acceptance rates at $\kappa=0.2705$ of the full independence test, Eq.~\eqref{eq:alpha-TV}, with the exact $\log w=-S-\log q$, and of the marginal test, Eq.~\eqref{eq:margtest}, with $\Phi=M$, for the proposals~\eqref{eq:qref}. Both are estimated from all pairs of 400 configurations generated with plain HMC and Wolff updates and 4000 proposals; the values of $\alpha_M$ agree with those measured in the chains. $\sigma^2$ is obtained from the interquartile range of $\log w$, $\sigma=(q_{75}-q_{25})/1.349$, because for a Gaussian proposal of an interacting target the distribution of $\log w$ has a long lower tail: proposals with large amplitudes have tiny weights from the quartic term. The last column gives the acceptance rate of the marginal test measured in the chains with the trained sampler that has the same reference process (Tables~\ref{tab:osq-samplers} and~\ref{tab:osq-diag}) as proposal, as the range over $\nref$; the full-test rates of these samplers are given in Tables~\ref{tab:osq-overlap} and~\ref{tab:osq-diag}.}
\label{tab:chainacc}
\end{table*}

\begin{figure*}[t]
\centering
\includegraphics[width=0.85\textwidth]{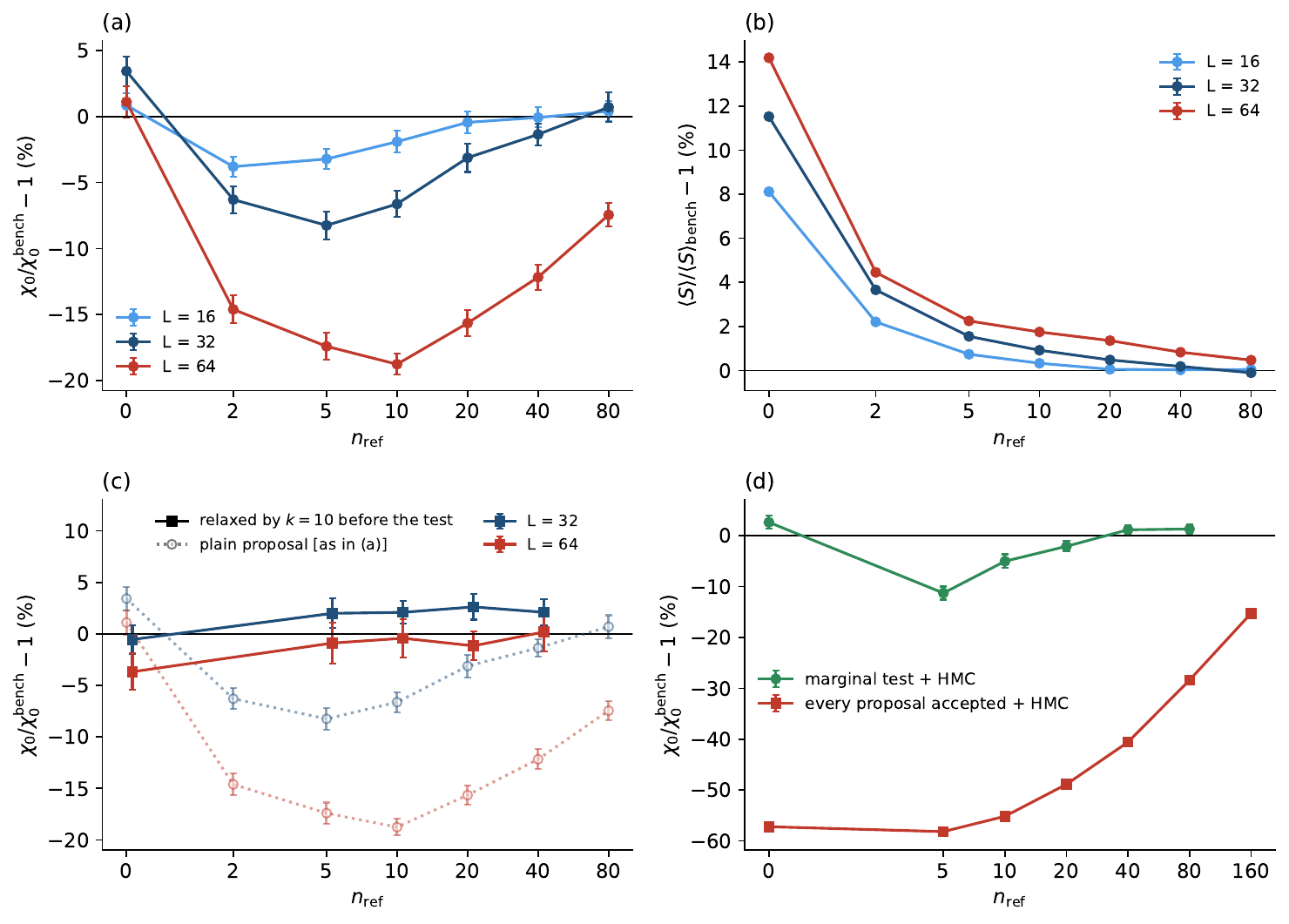}
\caption{Composite chain at $\kappa=0.2705$ with the reference process~\eqref{eq:qref} as proposal, marginal test on $M$  with $p_M$ from long Monte Carlo ensembles. (a) Relative deviation of $\chi_0$ from its benchmark value against the number $\nref$ of plain HMC trajectories per cycle, reference of Table~\ref{tab:osq-samplers}. (b) The same for $\langle S\rangle/V$, same runs. (c)  $\chi_0$ for proposals  relaxed by $k=10$ HMC trajectories before the test, Eq.~\eqref{eq:qk} (filled squares), compared with (a) (open circles). (d)  $\chi_0$ at  $L=32$  for the  reference of sampler A (Table~\ref{tab:osq-diag}): marginal test followed by HMC, and every proposal accepted followed by HMC.}
\label{fig:chain2705}
\end{figure*}

\paragraph{Ultraviolet relaxation and the zero mode.} The second cost is the number of trajectories needed to remove the ultraviolet mismatch. Figure~\ref{fig:chain2705} shows the chain with the reference process as proposal. Its ultraviolet error is removed quickly [Fig.~\ref{fig:chain2705}(b)]: within ten trajectories the excess of $\langle S\rangle/V$ drops from $8$--$14\%$ to below $2\%$ at all three volumes. The action density is dominated by the short-distance modes, which relax within a few trajectories, as in the free field, where each trajectory reduces the error of a mode with $\omega_k\bar t\gg1$ by about a factor of two [Proposition~\ref{prop:exact}(iv)]. Removing most of this error thus takes no more trajectories as $V$ grows over the range studied, despite its effect on the full test. The small remainder, largest at $L=64$, decays on the slower time scale of the infrared modes discussed below.

\begin{figure*}[t]
\centering
\includegraphics{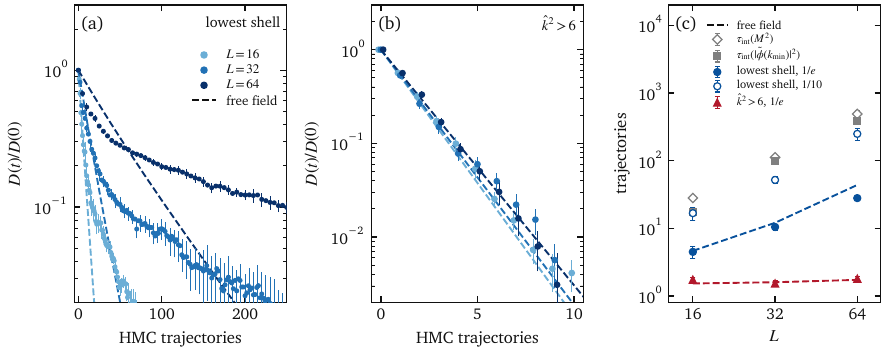}
\caption{Relaxation under plain HMC at $\kappa=0.2705$. Two-sided quench: in 512 configurations of the benchmark ensemble the Fourier amplitudes of a set of modes are multiplied by $\sqrt{1\pm d}$, and $D(t)$ is the difference between the two starts of the mean of $|\tilde\phi(k)|^2/V$ over the set after $t$ HMC trajectories. (a) Lowest shell, $1\le|n|<2$, $d=0.5$; at $L=32$, $d=0.2$ gives the same curve. (b) Ultraviolet modes, $\hat k^2>6$, $d=0.1$; since the change of all these amplitudes is extensive, a larger $d$ would change the HMC acceptance rates of the two starts. Dashed: free-field prediction for every mode of the set with $\omega_k^2=1/G(k)$, the measured $G(k)$ and HMC acceptance rate, and the leapfrog step and distribution of trajectory lengths of the simulation, weighted with the initial amplitude of each mode; Proposition~\ref{prop:exact}(iv) is its limit $\epsilon\to0$. (c) Times at which $D(t)/D(0)$ falls to $1/e$ (filled circles and triangles) and to $1/10$ (open circles), the $1/e$ time of the free-field prediction (dashed), and the equilibrium integrated autocorrelation times of $M^2$ and $|\tilde\phi(k_{\min})|^2$ under plain HMC (gray; 16 independent chains of $4$--$6\times10^4$ trajectories). Errors of the quench from a jackknife over the groups of configurations that share the random trajectory lengths, errors of $\tau_{\rm int}$ from a jackknife over chains.}
\label{fig:relax}
\end{figure*}

The zero mode, however, does not stay exact during this relaxation [Fig.~\ref{fig:chain2705}(a)]. At $\nref=0$ the distribution of $M$ is exact by Eq.~\eqref{eq:stationary}, up to the accuracy of the estimate of $p_M$, but for $\nref>0$, $\chi_0$ first decreases and then relaxes back toward the benchmark value. The dip deepens with $L$: at $L=64$, $\chi_0$ falls by about a fifth and is still $7.5(9)\%$ below the benchmark after 80 trajectories. This effect, absent in the free field, comes from the quartic term, which couples the zero mode to the short-distance fluctuations. Write $\phi_x=M+\delta\phi_x$ with $\sum_x\delta\phi_x=0$. Then
\begin{equation}\label{eq:Mcoupling}
\begin{split}
\sum_x\phi_x^4={}&VM^4+6M^2\sum_x\delta\phi_x^2\\
&+4M\sum_x\delta\phi_x^3+\sum_x\delta\phi_x^4,
\end{split}
\end{equation}
where the term $4M^3\sum_x\delta\phi_x$ is absent because $\sum_x\delta\phi_x=0$. The zero mode thus feels the short-distance fluctuations through the term $6\lambda M^2\sum_x\delta\phi_x^2$, which in the Hartree (tadpole) approximation is an extra mass for $M$ proportional to $\langle\delta\phi_x^2\rangle$. A proposal with too much short-distance fluctuation gives $M$ too large a mass. While the first trajectories remove the excess, HMC also moves $M$, now in the wrong effective potential, toward smaller $|M|$. After that $M$ relaxes back on the zero-mode time scale, the slowest near criticality [Fig.~\ref{fig:relax}(c)]. Proposition~\ref{prop:exact}(i) is therefore not guaranteed with interactions: for the proposals studied here, the stationary zero-mode distribution is exact up to marginal-density errors at $\nref=0$, deviates from the target at intermediate $\nref$, and relaxes toward it at large $\nref$. An ultraviolet error becomes expensive to remove after the test when its removal disturbs the slowest mode.

The remedy is to reverse the two steps. The marginal test needs only the marginal $q_\Phi$ of the proposal mechanism, not $q(U)$, and any procedure that does not depend on the current state of the chain is an allowed proposal. We may apply $k$ HMC trajectories at the target to a generator sample before the test, giving density
\begin{equation}\label{eq:qk}
q_k(U)=\int dU_0\,q(U_0)\,K_{\rm loc}^k(U_0,U),
\end{equation}
which is unknown, but its marginal $q_{k,\Phi}$ can be estimated from relaxed proposals, as for $p_\Phi$. With exact marginal densities, the test has stationary distribution $p_\Phi(\Phi)\,q_k(U|\Phi)$, by Appendix~\ref{app:stat} with $q\to q_k$. Since $K_{\rm loc}$ leaves $p$ invariant, $q_k$ is no further from $p$ than $q$ in this divergence: $D_{\rm KL}(p\|q_k)\le D_{\rm KL}(p\|q)$ by the data-processing inequality~\cite{Cover:2006}, applied to the same kernel acting on $p$ and on $q$. The zero mode is now pulled \emph{before} the test, which corrects it. With $k=10$ the dip disappears [Fig.~\ref{fig:chain2705}(c)]: at $L=32$ and $64$, $\chi_0$ agrees with the benchmark value within about two standard deviations for all $\nref$, and the acceptance rate stays at $0.81$ and $0.72$. The ultraviolet error is then removed at the cost of $k$ additional trajectories per cycle, without a resolved shift of the zero mode in these runs.

\begin{figure*}[t]
\centering
\includegraphics{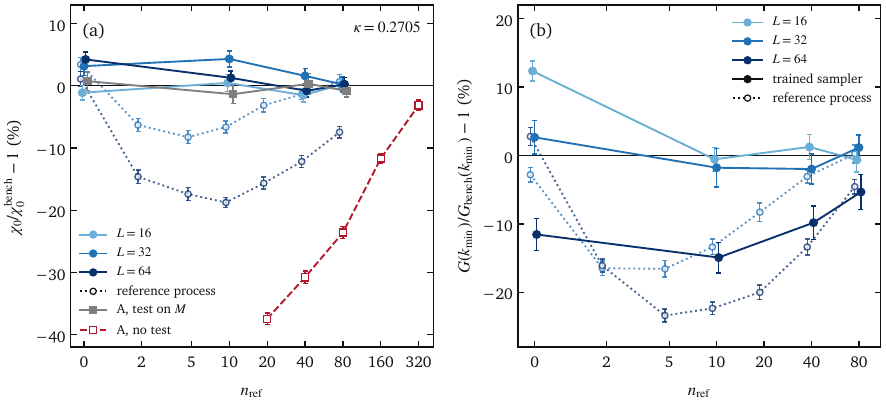}
\caption{Composite chain at $\kappa=0.2705$ with the trained samplers as proposals, marginal test on $M$  with $p_M$ from long Monte Carlo ensembles, followed by $\nref$ plain HMC trajectories. (a) Relative deviation of $\chi_0$ from its benchmark value against $\nref$ for the samplers of Table~\ref{tab:osq-samplers} (filled circles), for their reference processes at $L=32$ and $64$ [open circles, dotted; the runs of Fig.~\ref{fig:chain2705}(a)], and for sampler A of Table~\ref{tab:osq-diag} with the marginal test (filled squares) and with every proposal accepted (open squares, dashed). (b) Relative deviation of the lowest nonzero mode, $G(k_{\min})$, for the samplers of Table~\ref{tab:osq-samplers} and their reference processes, same runs as in (a). Points at the same $\nref$ are drawn slightly displaced.}
\label{fig:chainosq}
\end{figure*}

With trained proposals [Fig.~\ref{fig:chainosq}(a)], $\chi_0$ shows no dip, and relaxing the proposals before the test does not reduce the deviations. Their ultraviolet is already accurate: at $\nref=0$ their $\langle S\rangle/V$ is within $1\%$ of the target at all three volumes, so local updates need not remove excess short-distance fluctuations that would pull the zero mode. The remaining deviations concern the width of $|M|$: at $L=32$ and $64$ the chain at small $\nref$ still has too few configurations with small $|M|$, like the samplers themselves (Sec.~\ref{sec:hmc-osq}), so that $\chi$ lies below the benchmark and $\chi_0$ a few percent above it. A test with exact marginal densities would remove this  discrepancy  at $\nref=0$, where $\chi$ is still $20(4)\%$ ($L=32$) and $10(3)\%$ ($L=64$) below the benchmark. Since the test is limited by the estimates of $p_M$ and $q_M$, the  discrepancy   most likely comes from  the kernel density estimate of $q_M$, which smooths out the lack of samples at small $|M|$: for the reference process, whose $q_M$ is known in closed form and whose $\chi$ also lies below the benchmark (Table~\ref{tab:osq-samplers}), the test leaves no such  discrepancy  at $\nref=0$.  After 80 HMC trajectories, $\chi$ is consistent with the benchmark  at $L=32$, whereas at $L=64$ the deviation of $\chi$ hardly decreases, like that of the lowest nonzero modes discussed below.

\paragraph{Infrared errors.} We next consider an infrared error. For the reference of sampler A at $L=32$, $\chi_0$ is less than half the benchmark value. With the marginal test, which accepts $0.40$--$0.47$ of the proposals, the chain reaches the benchmark $\chi_0$ within errors after 40 trajectories [Fig.~\ref{fig:chain2705}(d)]; the dip at $\nref=5$ is the pull of the zero mode described above. Accepting every proposal and relying on the local updates alone, as rethermalization does in inverse blocking~\cite{Hasenfratz:2026,Hasenfratz:2026b}  but without its matched coarse ensemble, leaves $\chi_0$ more than $10\%$ below the benchmark even after 160 trajectories. Sampler A itself, whose zero mode is also far from the benchmark (Table~\ref{tab:osq-diag}), shows the same effect without the dip [Fig.~\ref{fig:chainosq}(a)]. With the marginal test, which accepts $0.54$--$0.56$ of its proposals, $\chi_0$ agrees with the benchmark value within errors for all $\nref$, already at $\nref=0$. Without the test, $\chi_0$ is still about $12\%$ below the benchmark after 160 trajectories, and about 300 trajectories bring it within a few percent of it. Unlike an ultraviolet error, an error in a relevant direction is removed only slowly by local updates, on the time scale of the zero mode, as also observed for a mistuned thermal coupling in inverse blocking~\cite{Hasenfratz:2026}, whereas an exact marginal test has the correct stationary distribution in that direction. Samplers B and C, whose  $\langle|M|\rangle$ and $\chi_0$  are close to the target, behave like the samplers of Table~\ref{tab:osq-samplers}; the largest deviation at small $\nref$ occurs for B, whose distribution of $|M|$ is the narrowest (Sec.~\ref{sec:phi4-ir}), as expected from the estimate of $q_M$.

\paragraph{Untested infrared modes.} Figure~\ref{fig:chainosq}(b) shows the deviation of the untested lowest nonzero mode, $G(k_{\min})$. For the trained samplers its  deviation is consistent with zero after  ten trajectories at $L=16$ and is within the statistical errors at all $\nref$ at $L=32$, whereas at $L=64$, where the sampler lies about $10\%$ below the benchmark for this mode, the error decreases only slowly with $\nref$. The reference process shows the same slow mode: the relaxation of its ultraviolet pulls $G(k_{\min})$ down in the same way as $\chi_0$, and even with relaxed proposals $G(k_{\min})$ is still $22(2)\%$ below the benchmark after 40 trajectories at $L=64$. The slow modes are thus those outside $\Phi$. Near criticality $\xi$ is comparable to $L$, so the lowest nonzero modes relax under local updates on a time scale similar to the zero mode. Such modes can be included in the collective variables, and their number does not grow with the volume: when $\xi$ is proportional to $L$ the shells with $\kh\lesssim1/\xi$ form a fixed set to include in $\Phi$; in the free-field picture, including it shortens the relaxation time by a constant factor. With $\Phi=(M,\log g_1)$, where $g_1$ is $|\tilde\phi(k)|^2/V$ averaged over the lowest nonzero momenta,  and with the joint target density of $M$ and $g_1$ as additional input,  the deviation of $G(k_{\min})$ at $L=64$ and $\nref=40$ is consistent with zero, both for the reference process and for the trained sampler with relaxed proposals (Table~\ref{tab:cost}); the test then accepts $0.54$--$0.60$ and $0.74$--$0.77$ of the proposals, respectively. For the reference process this is a clear improvement over the test on $M$ alone. For the trained sampler the statistical error is too large to establish an improvement over the $-9.8(25)\%$ obtained with $\Phi=M$, and without relaxing the proposals the deviation is not significantly reduced. The size of $\Phi$ is limited by the accuracy of $p_\Phi/q_\Phi$ in higher dimensions: with the kernel density estimates used here, $G(k_{\min})$ still deviates by $-17(4)\%$ for the reference process at $\nref=0$, where the test alone would have the exact stationary distribution of $g_1$ if the marginal densities were known.

\begin{table*}[t]
\centering\small
\setlength{\tabcolsep}{3pt}
\begin{tabular}{clcccccc}
\toprule
 & & & \multicolumn{3}{c}{ cost (HMC trajectories) } & \multicolumn{2}{c}{ deviation (\%) }\\
\cmidrule(lr){4-6}\cmidrule(lr){7-8}
$L$ &  update  & $\Phi$ & \parbox[b]{20mm}{\centering per cycle \\ $k+\nref+c_{\rm gen}$ } & \parbox[b]{19mm}{\centering cycles \\ $\times\,2\tau_{\rm int}(M^2)$ } & \parbox[b]{10mm}{\centering total \\ $=C_{M^2}$ } & $\chi_0$ & $G(k_{\min})$\\
\midrule
16 & plain HMC & -- &  1  &  56  & 56 & -- & --\\
 &  trained  & $M$ &  $0+40+35$  &  1.1  & 85 & $-1.5(11)$ & $+1.3(18)$\\
\midrule
32 & plain HMC & -- &  1  &  223  & 223 & -- & --\\
 &  reference  & $M$ &  $0+40+0$  &  1.7  & 68 & $-1.4(8)$ & $-3.1(10)$\\
 &  reference, relaxed  & $M$ &  $10+40+0$  &  1.4  & 71 & $+2.1(13)$ & $-5.7(19)$\\
 &  reference, relaxed  &  $(M,\log g_1)$  &  $10+10+0$  &  1.9  & 38 & $+0.9(15)$ & $-4.6(31)$\\
 &  trained  & $M$ &  $0+40+45$  &  1.3  & 113 & $+1.6(12)$ & $-2.0(22)$\\
\midrule
64 & plain HMC & -- &  1  &  980  & 980 & -- & --\\
 &  reference  & $M$ &  $0+80+0$  &  1.8  & 141 & $-7.5(9)$ & $-4.6(11)$\\
 &  reference, relaxed  & $M$ &  $10+10+0$  &  2.6  & 52 & $-0.4(19)$ & $-30(2)$\\
 &  reference, relaxed  &  $(M,\log g_1)$  &  $10+40+0$  &  4.3  & 217 & $-1.3(20)$ & $+2.4(35)$\\
 &  trained  & $M$ &  $0+40+29$  &  1.6  & 110 & $-0.8(10)$ & $-9.8(25)$\\
 &  trained, relaxed  &  $(M,\log g_1)$  &  $10+40+29$  &  2.1  & 164 & $+1.9(17)$ & $-2.6(44)$\\
\bottomrule
\end{tabular}
\caption{ Cost of an effective independent sample of $M^2$ and remaining infrared deviations at $\kappa=0.2705$. The cost per cycle, in HMC trajectories, is $k+\nref+c_{\rm gen}$: $k=10$ trajectories applied to relaxed proposals before the test, $\nref$ trajectories after it, and the generator time $c_{\rm gen}$ converted into trajectories. For plain HMC a cycle is one trajectory. Multiplied by the number of cycles per independent sample, $2\tau_{\rm int}(M^2)$, it gives $C_{M^2}$. Reference proposals come from the reference process, Eq.~\eqref{eq:qref}, whose generation cost is omitted; trained proposals come from the OSQ samplers of Table~\ref{tab:osq-samplers}. Deviations are relative to the benchmark; plain-HMC autocorrelation times are from Fig.~\ref{fig:relax}(c). Costs exclude training and the determination of $p_\Phi$. On RIKYU, generation takes $0.70$, $3.2$ and $8.3$~ms per sample at $L=16$, $32$ and $64$ on an NVIDIA GB200 GPU (batches of 2048, 250 Euler--Maruyama steps), and one HMC trajectory takes $0.020$, $0.072$ and $0.29$~ms on one Arm Neoverse-V2 core (compiled Numba implementation); this conversion compares batched GPU generation with single-core CPU HMC and favors the generator. }
\label{tab:cost}
\end{table*}

The time scales of the local updates can be measured directly, without the test, in a two-sided quench under plain HMC (Fig.~\ref{fig:relax}). In configurations of the benchmark ensembles the Fourier amplitudes of a set of modes are multiplied by $\sqrt{1+d}$ or by $\sqrt{1-d}$, both starts are evolved with the same random numbers, and their difference $D(t)$ in mean mode power is followed over $t$ trajectories; $D(t)$ tends to zero, and the equilibrium value drops out. For the ultraviolet modes, $\hat k^2>6$, $D(t)$ decays in the same way at all three volumes, with a time constant of about two trajectories [Fig.~\ref{fig:relax}(b)]. The decay agrees with the free-field relaxation of Proposition~\ref{prop:exact}(iv) for modes with $\omega_k^2=1/G(k)$ at the measured HMC acceptance rate: short-distance modes relax like free modes with frequencies set by the measured propagator. The ultraviolet relaxation time therefore does not grow with $V$ for $L=16$--$64$, consistent with the free-field prediction of no further volume dependence at fixed acceptance.

The lowest shell, $1\le|n|<2$ with $k=2\pi n/L$, behaves differently [Fig.~\ref{fig:relax}(a)]. At $L=16$ the free-field prediction describes its decay to $1/e$, but at larger $L$ it becomes faster initially and slower later: at $L=64$, $D(t)$ falls to $1/e$ of its initial value after $28(2)$ trajectories, against 44 in the free field, but to a tenth only after about 250, against 110 [Fig.~\ref{fig:relax}(c)]. The late tail follows the zero mode. Raising the amplitudes of the lowest modes lowers $V\langle M^2\rangle$ through the coupling~\eqref{eq:Mcoupling}, by up to about a tenth of its value, and this difference relaxes over several hundred trajectories at $L=64$; for $t\gtrsim200$ the tail of $D(t)$ is proportional to it. Equilibrium autocorrelations show the same slowing [gray points in Fig.~\ref{fig:relax}(c)]: the integrated autocorrelation times of $M^2$ and of $|\tilde\phi(k_{\min})|^2$ under plain HMC, which describe a different correlation function, grow from $28(1)$ and $18(1)$ trajectories at $L=16$ to $490(60)$ and $380(40)$ at $L=64$, like $L^{2.06(9)}$ and $L^{2.20(9)}$. This is the critical slowing down of local updates when $\xi$ is comparable to $L$, consistent with the free-field scaling of Eq.~\eqref{eq:taum} at $\xi\propto L$: with $\omega_{\min}\simeq1/\xi$ the bound gives $\tau_m\simeq\xi^2/2\bar t^{\,2}$, that is, $z=2$. At fixed $\xi$ the same bound is independent of $L$, and the measured times at fixed $\xi$  show no resolved growth  once $L$ is several times $2\pi\xi$ (Appendix~\ref{app:fixedxi}); no exponential growth with $V$ is evident over the volumes studied.

\paragraph{ Decorrelation cost and residual bias. }
 Table~\ref{tab:cost} compares the cost of refreshing the zero mode with the errors that remain in the sampled configurations. The target marginals are supplied by the long ensembles used above. Since these already determine $\langle M^2\rangle$, we use $M^2$ here to measure chain decorrelation; the deviation columns assess the accuracy of the same chain. 

 One cycle generates a proposal, applies $k=0$ or $10$ plain HMC trajectories to the proposal before the marginal test, and then applies $\nref$ trajectories to the chain after accepting or rejecting the proposal. These HMC updates allow $M$ to change. If generation costs $c_{\rm gen}$ in HMC trajectory units, the cost per effective independent measurement of $M^2$ is 
\begin{equation*}
 C_{M^2}=2\tau_{\rm int}(M^2)\bigl(\nref+k+c_{\rm gen}\bigr), 
\end{equation*}
 where $\tau_{\rm int}$ is measured in cycles. Table~\ref{tab:cost} lists both factors and their product. For plain HMC, $C_{M^2}=2\tau_{\rm int}(M^2)$ with $\tau_{\rm int}$ measured in trajectories. 

 For example, at $L=64$ the trained proposal with $\Phi=M$ costs $40+29=69$ trajectory units per cycle. An autocorrelation factor $2\tau_{\rm int}\simeq1.6$ gives $C_{M^2}=110$, compared with 980 for plain HMC. This faster decorrelation coexists with deviations of $-9.8(25)\%$ in $G(k_{\min})$ and $-14.3(31)\%$ in $\chi$, despite the small deviation in $\chi_0$. Including $g_1$ in the test and relaxing proposals by $k=10$ trajectories raises the cost to 164; $G(k_{\min})$ then differs by $-2.6(44)\%$, consistent with zero, although the improvement over the $M$-only test is not statistically resolved. The cost must therefore be read together with the remaining deviations: these settings are not compared at equal accuracy. Moreover, $C_{M^2}$ excludes training and the determination of $p_\Phi$, and converts batched GPU generation time into single-core CPU trajectories, which favors the generator.  Plain HMC is a relevant baseline for gauge theories; for $\phi^4$ itself cluster algorithms can be substantially faster.

The cost per trajectory still grows with $V$. For leapfrog HMC at fixed trajectory length and finite correlation length, the usual extensive-error estimate gives $\langle\Delta H^2\rangle\propto V\epsilon^4$, so that a fixed acceptance rate requires $\epsilon\propto V^{-1/4}$ and a cost per trajectory $\propto V^{5/4}$~\cite{Creutz:1988,Gupta:1988}; this is the same for plain HMC. Ultraviolet correction therefore needs a number of trajectories independent of $V$ for $L=16$--$64$, with a per-trajectory cost estimated as proportional to $V^{5/4}$, while infrared modes outside $\Phi$ add a local relaxation time growing as a power of $L$ near criticality [Fig.~\ref{fig:relax}(c)]. These estimates, for a given $p_\Phi$,  suggest polynomial growth with $V$, in contrast to the exponential cost associated with a fixed extensive mismatch in global reweighting or independence testing.

The results above take $p_M$ from the long ensembles of each implementation. If $p_M$ is instead estimated from a short auxiliary run of 400 or 4000 configurations in total (about 30 and 300 independent ones), the deviation of $\chi_0$ at $L=32$ and $\nref=40$ with the reference process grows by $1$--$2\%$. This additional deviation is a systematic error of the method (Sec.~\ref{sec:fix}).

 Appendix~\ref{app:pphi} examines an alternative determination of $p_M$ from constrained simulations of the action. The resulting $\hat p_M$ is consistent with the benchmark at $L=32$ and $64$, whereas, with $N=10240$ samples, the samplers' own path weights give a reliable $p_M$ only at $L=16$. 

In summary, the ultraviolet configuration mismatch, which contributes to the exponential decrease of the ESS and global acceptance with $V$, is  largely  removed by exact local updates at a cost consistent with nonexponential growth over the range studied: the marginal test accepts at a rate of order one, and at every volume studied the ultraviolet modes relax with a time constant of about two trajectories and about ten trajectories remove most of the ultraviolet error of the proposals. Two refinements of the free-field picture are needed. First, with a proposal whose ultraviolet is wrong, short-distance relaxation pulls the zero mode, an effect reduced by proposal relaxation before the test; for the trained samplers, whose ultraviolet is already accurate, no pull is resolved. Second, when $\xi$ is comparable to $L$, the lowest nonzero modes relax slowly as well, and they have to be reproduced by the sampler or included in $\Phi$. In both cases the expensive correction comes from the infrared, where the sampler and test must be accurate.

\section{Discussion}\label{sec:discussion}

\subsection{Scale-resolved assessment of samplers}
The first question in the introduction motivates diagnostics at the scales probed by the observable. A global ESS does not locate the mismatch and can itself be overestimated when rare large weights are missed (Secs.~\ref{sec:estimators} and~\ref{sec:hmc-osq}). We therefore propose reporting the ESS together with $\sigma^2/V$, the overlap flow $D_2(\Lambda)$ or its Gaussian proxy from the propagators, the overlap of the zero mode and of sectors, and the global and marginal acceptance rates against $V$. For OSQ, the log-weight diagnostics include fluctuations of the generating paths, whereas the marginal and propagator diagnostics probe the final fields. The latter need only unweighted samples and reference information on quantities such as $M$, $VM^2$ and low Fourier modes, so they also apply to samplers without a closed-form density~\cite{Tan:2026}. Schwinger--Dyson identities provide a complementary check of a learned score without sampling from it~\cite{Schubring:2026}.

In the free-field continuum comparison, overlap loss  can come  from a lattice artifact whose effect on fixed-momentum observables vanishes (Sec.~\ref{sec:free}). Symmetry provides another way to constrain the infrared error. A weak $\mathbb{Z}_2$-breaking perturbation $h\sum_x\sigma_x$ in the two-dimensional Ising class gives $D_2\sim Vh^2\xi^{7/4}$; exact $\mathbb{Z}_2$ symmetry excludes this term. Approximate symmetry, for example through symmetrized training data, can leave a residual $h$ that biases long-distance observables and suppresses the ESS exponentially in $V$.

\subsection{Implications for sampler design}
For the second question, the relaxation bound translates into a requirement on the proposal's slow modes. In the free field, an untested mode's residual variance bias equals its proposal bias times the mean survival factor under local updates. Exponential relaxation suggests $\nref\simeq\tau(\xi)\log(b_0/\delta)$ trajectories to reduce an initial relative bias $b_0$ to $\delta$. In the free-field analysis, $\tau$ is controlled by the slowest mode outside $\Phi$ and saturates with $L$ at fixed $\xi$ once $L\gg\xi$ (Secs.~\ref{sec:fix} and~\ref{sec:exact}). Near criticality, it grows with $L$: fresh proposals can  decorrelate $M^2$ within a few cycles  while leaving a residual bias in the untested modes  (Table~\ref{tab:cost}) . Improving those modes in the generator reduces the required relaxation, without eliminating the critical slowing of the local kernel. The interacting tests illustrate this distinction (Sec.~\ref{sec:phi4chain}).

A global objective such as Eq.~\eqref{eq:osq-loss} can be dominated by the ultraviolet, making a separate parametrization of the infrared useful. In the OSQ samplers, the two reference parameters  mainly  control the zero mode, leaving the network, which acts on all modes,  to learn the remaining drift. Sampler B shows why training must also be assessed through infrared diagnostics: its magnetization distribution remains inaccurate  while its log-weight variance already agrees within $1\%$ with that of sampler C, the same run after 5000 steps  (Sec.~\ref{sec:phi4-ir}).

Alternatively, coarse-grained data or physical priors can supply the infrared while the generator supplies the ultraviolet~\cite{Marchand:2022,Bauer:2024,Singha:2026,Hasenfratz:2026};  such input could provide $p_\Phi$ for the composite chain (Sec.~\ref{sec:fix}).  Intermediate scales could be treated with proposals on blocks of fixed size $\ell^d$, conditioned on their boundary, with local acceptance of order $e^{-c\ell^d}$ rather than $e^{-cV}$~\cite{Finkenrath:2022,Finkenrath:2024}; we have not tested this extension.  A simpler exact alternative is a uniform shift $\phi_x\to\phi_x+\delta$, with $\delta$ drawn from a symmetric distribution and acceptance $\min[1,e^{-\Delta S}]$ computed from the full action. Interleaving these moves with exact local updates preserves the target distribution without requiring $p_M$. In the free field the zero mode can also be drawn directly from its Gaussian target; with interactions, a zero-mode update at fixed nonzero modes samples a conditional distribution, not $p_M$. We have not compared the efficiency of these alternatives with the composite chain. A comparison at equal accuracy would need to include the cost of obtaining $p_M$ and the relaxation of the other long-wavelength modes. 

This separation of scales also underlies multigrid Monte Carlo~\cite{Goodman:1986}, multiscale equilibration~\cite{Endres:2015} and inverse blocking~\cite{Hasenfratz:2026,Hasenfratz:2026b}. Our analysis identifies a marginal test as a way to correct errors in the chosen coarse variables: with exact density ratios, the test alone has their target distribution at stationarity. The remaining conditional mismatch is left to rethermalization, which can in turn perturb the tested variables. The composite chain approaches the full target as $\nref$ grows, assuming convergence of the local kernel; it is generally biased at finite $\nref$.  Inverse blocking supplies a joint infrared distribution implicitly through a matched coarse ensemble, whereas the marginal test needs the target distribution of selected variables explicitly. The two could be combined: the matched coarse ensemble would supply the infrared of the proposals, and a marginal test on a few coarse observables, with $p_\Phi$ computed from the action, would correct the residual mismatch in directions that rethermalization repairs slowly. This hybrid has not been tested; the computation of $\hat p_M$ in Appendix~\ref{app:pphi} shows only that such input can be obtained for $\Phi=M$ at $L=32$ and $64$. 

 The trained samplers illustrate the use of such marginal information.   For observables that involve $M$ together with other modes,  reweighting by $p_M/q_M$  corrects only the distribution of $M$, not the model's conditional distribution at fixed $M$, and  has an asymptotic ESS fraction set by $D_2(p_M\|q_M)$: the fixed-reference values $0.03$--$0.21$ give $\ESS/N=0.81$--$0.97$. This could also reduce the zero-mode bias reported in Ref.~\cite{Tan:2026}. For the composite chain, the main outstanding issues are the relaxation of untested infrared modes on larger near-critical lattices and the accuracy of $p_\Phi/q_\Phi$. At fixed correlation length, $\xi\simeq5$,  the  quench and equilibrium  relaxation times  of the lowest modes show no growth between $L=128$ and $256$  (Appendix~\ref{app:fixedxi}). Density-estimation errors already limit the test for $\Phi=M$, and more so for $\Phi=(M,\log g_1)$ (Sec.~\ref{sec:phi4chain}).

In gauge theories possible choices of $\Phi$ include the topological charge or the Polyakov loop. A marginal test can correct sector weights only if the proposal reaches all target sectors (Eq.~\eqref{eq:sector}). Sector-resolved samplers, which determine relative sector weights from restricted partition functions~\cite{Singha:2026b}, address sector-weight calibration within the generator.

 In Sec.~\ref{sec:phi4} we took $p_\Phi$ from long Monte Carlo ensembles and, for $\Phi=M$ at $L=32$ and $64$, estimated it from constrained simulations of the action, which used about half as many HMC trajectories as plain HMC is estimated to need for the same statistical error of $\chi$, not counting training, generation and the benchmark calibration of the starting samples (Appendix~\ref{app:pphi}). Whether such input can be obtained at controlled cost on much larger volumes has not been tested. For $L\gg\xi$ the distribution of the volume average of a local quantity, such as $M$, approaches a Gaussian, or a pair of Gaussians in the broken phase; for $\xi$ comparable to $L$ it approaches a finite-size scaling form. Such forms, with parameters determined at moderate volume, could reduce the calibration needed at each volume, but their accuracy, in particular in the tails, and their range of validity remain to be tested. Even for a fixed number of tested variables, determining their joint density accurately remains a separate task, which becomes harder as their number grows, as noted above for $\Phi=(M,\log g_1)$. For the topological charge, $p(Q)$ at large volume is, to leading order, a Gaussian of variance $\chi_tV$ for $|Q|\ll\chi_tV$, where $\chi_t$ is the topological susceptibility~\cite{Aoki:2007}; the corrections involve higher cumulants, which are nonzero in SU(3) Yang--Mills theory~\cite{Ce:2015}. Conventional simulations at fixed topology give the sector weights only indirectly, for example through $\chi_t$ extracted from correlators within one sector~\cite{Aoki:2007}; this large-volume form would reduce sector calibration to a few intensive numbers, $\chi_t$ and the leading higher cumulants, whose determination near the continuum limit remains difficult. 

\section{Conclusions}\label{sec:concl}

The first question is answered by resolving the overlap by scale. The asymptotic ESS fraction, $e^{-D_2(p\|q)}$, measures the aggregate mismatch, whereas a long-distance observable probes the marginal distribution of the variables on which it depends. Errors spread over many short-distance modes can therefore suppress the global ESS without appreciably biasing that observable. The free-field comparison makes this distinction explicit: two samplers with comparable ESS give a $32\%$ bias and no bias in the susceptibility. The overlap flow $D_2(\Lambda)$ distinguishes their errors as short-distance modes are removed.

The trained OSQ samplers provide interacting evidence over the volumes studied.  With reference parameters fixed from the benchmark, they  reproduce $\langle|M|\rangle$ and $\langle M^2\rangle$ to a few percent while the path log-weight variance grows approximately with volume, much of it apparently from path fluctuations rather than endpoint mismatch.  At $L=32$, changing  the reference parameters substantially changes the infrared observables while leaving this variance nearly unchanged.

 For the second question, the marginal test shows how a given target marginal can be used in full-configuration sampling without the exponential penalty of global overlap for the observables studied here. Observables of $M$ alone already follow from $p_M$; the chain is useful for other infrared modes and for observables coupling them to $M$, subject to the remaining conditional error.  The composite chain generally has a biased stationary distribution at finite relaxation length and approaches the full target as the local kernel converges. With exact marginal density ratios, the marginal test accepts at least as often as the full independence test at their respective stationary distributions. The free-field solution bounds the number of local updates needed for a fixed accuracy of the mode variances independently of volume, at fixed correlation length and uniformly bounded relative errors in the proposal mode variances. In $\phi^4$ theory  near criticality, the marginal acceptance remains of order one and about ten trajectories remove most of the ultraviolet mismatch at all volumes studied, $L=16$--$64$ .

The remaining limitations concern infrared relaxation and marginal-density estimation.  Interactions can transfer a proposal's ultraviolet error into the tested zero mode during relaxation, so the ultraviolet is best relaxed before the infrared is corrected, as with the relaxed proposals here, or with the tested variables held fixed.  Untested low modes can remain slow near criticality, and inaccurate marginal ratios leave a systematic error.  In $\phi^4$ theory at $L=32$ and $64$, integrating the mean force at fixed $M$ gives a marginal $\hat p_M$ consistent with the benchmark (Appendix~\ref{app:pphi}), although both kinds of starting samples use benchmark-calibrated reference parameters; drawing $M$ from it avoids estimating $q_M$, while the uncertainty of $\hat p_M$ and the error of the untested low modes remain. 

 More generally, when the untested modes carry an extensive mismatch, generative proposals do not remove the need to determine the target infrared distribution: their own weights give it only at high cost, so it has to be supplied separately, for example by constrained simulations or by an RG-matched coarse ensemble as in inverse blocking. Obtaining joint densities of several variables, topological sector weights, and such input on much larger volumes at controlled cost remains open; our results provide a scale-resolved assessment and a conditional use of infrared information, not a solution of the volume problem. 

\begin{acknowledgments}
We thank Gert Aarts, Yuto Ashida, Anna Hasenfratz, Gurtej Kanwar, Jan Pawlowski and Phiala Shanahan for valuable discussions, and Yang-yang Tan for providing the benchmark ensembles and insightful discussions. We thank ECT$^*$ for support at the Workshop ``Bridging machine learning and statistical field theory through generative and diffusion processes'', during which this work has been developed. We also thank the members of the DM-QFT Collaboration and the DEEP-IN WG at RIKEN-iTHEMS for discussions. We acknowledge collaboration with the Google DeepMind team. This research used (in part) computational resources of the supercomputer RIKYU provided by the RIKEN Center for Computational Science.
This work was supported by JST ARiSE Program Japan Grant Number JPMJAR2633. S.I.\ is supported by JSPS KAKENHI Grant-in-Aid for Scientific Research (B), Grant No.\ 24K00634. L.W.\ is also supported by the JST-BOOST Grant No.\ JPMJBY24H9, the RIKEN-TRIP initiative (RIKEN-Quantum), and JSPS KAKENHI Grant No.\ 25H01560.
\end{acknowledgments}

\section*{Use of AI tools}
Claude Opus 5.5/Fable 5.1 (Anthropic) and ChatGPT 6.0 (OpenAI) assisted substantially with numerical cross-checks and with polishing the manuscript. All derivations, results, and statements were verified by the authors, who take full responsibility for the content.

\section*{Data availability}
The code of the OSQ samplers, on which our samplers are based, is publicly available at \url{https://github.com/DM-QFT/OSQ-Scalar}~\cite{Wang:OSQ}. The scripts used for the analysis in this work, including the training with a trainable reference, are available from the authors upon request. The benchmark ensembles are publicly available at \url{https://huggingface.co/datasets/phi4-generative/phi4-critical-configs}~\cite{Tan:2026data}.

\appendix

\section{\texorpdfstring{ Derivations }{Derivations}}\label{app:deriv}

\subsection{R\'enyi divergence and coarse graining}\label{app:ess}\label{app:dpi}\label{app:gibbs}

\paragraph{Monotonicity of the overlap flow.} Let $R(U)$ be a coarse variable (for example the modes with $\kh\le\Lambda$) and $Rp$, $Rq$ its densities. With $w=p/q$,
\begin{align*}
(Rp)(y)&=\int dU\,p(U)\,\delta(R(U)-y)\\
&=\int dU\,q(U)\,w(U)\,\delta(R(U)-y)\\
&=(Rq)(y)\,\Eq[w\,|\,R=y].
\end{align*}
Hence $Rp/Rq=\Eq[w|R]$. By Jensen's inequality, $(\Eq[w|R])^2\le\Eq[w^2|R]$, so
\begin{align*}
\int dy\,(Rq)\Big(\frac{Rp}{Rq}\Big)^2&=\Eq\Big[\big(\Eq[w|R]\big)^2\Big]\\
&\le\Eq\Big[\Eq[w^2|R]\Big]=\Eq[w^2].
\end{align*}
Taking the logarithm gives $D_2(Rp\|Rq)\le D_2(p\|q)$. If $R'$ is a function of $R$, the same argument applied to $R$ gives $D_2(R'p\|R'q)\le D_2(Rp\|Rq)$, which is the monotonicity of the flow.

\paragraph{Gibbs family.} With $p=e^{-S}/Z(0)$ and $q=e^{-S-cA}/Z(c)$,\[\frac{p^2}{q}=e^{-S+cA}\frac{Z(c)}{Z(0)^2},\qquad \int d\phi\,\frac{p^2}{q}=\frac{Z(-c)Z(c)}{Z(0)^2}.\]  Since $\log Z(c)=\log Z(0)+\sum_{n\ge1}(-c)^n\kappa_n/n!$, with $\kappa_n$ the cumulants of $A$ under $p$, the odd terms cancel and $D_2=2\sum_{n\,\rm even}c^n\kappa_n/n!=c^2\kappa_2+c^4\kappa_4/12+O(c^6)$, where $\kappa_2=\langle A^2\rangle_{\rm conn}=V\chi_O$. 

\subsection{Independence and marginal tests}\label{app:tests}\label{app:acc}\label{app:stat}

\paragraph{Acceptance rate.} With $x\sim p$, $y\sim q$ independent,
\begin{align*}
\alpha&=\int dx\,dy\,p(x)q(y)\min\Big[1,\frac{p(y)q(x)}{q(y)p(x)}\Big]\\
&=\int dx\,dy\min[p(x)q(y),q(x)p(y)],
\end{align*}
and $\int\min(f,g)=1-\frac12\int|f-g|=1-\TV(f,g)$ for probability densities. Using $\min(f,g)\le\sqrt{fg}$ gives $\alpha\le(\int\sqrt{pq})^2=e^{-D_{1/2}}$. For the marginal test the same calculation on the $\Phi$ space gives $\alpha_\Phi$. The map $(x,y)\mapsto(\Phi(x),\Phi(y))$ sends $p\otimes q$ to $p_\Phi\otimes q_\Phi$ and $q\otimes p$ to $q_\Phi\otimes p_\Phi$, and $\frac12\int|Kf-Kg|\le\frac12\int|f-g|$ for any Markov kernel $K$ (triangle inequality), so $\alpha_\Phi\ge\alpha$.

For Gaussian $\ell=\log w$ (mean $-\sigma^2/2$ under $q$, $+\sigma^2/2$ under $p$), $Z=\ell(y)-\ell(x)$ has mean $-\sigma^2$ and variance $2\sigma^2$, and $\alpha=P(Z>0)+\mathbb{E}[e^Z;Z<0]=2\Phi_{\rm N}(-\sigma/\sqrt2)$.

\paragraph{Stationary distribution.} With $r=p_\Phi/q_\Phi$ and $T(U\to U')=q(U')\min[1,r(\Phi')/r(\Phi)]$, the density $\pi(U)=q(U)r(\Phi)=\allowbreak p_\Phi(\Phi)q(U|\Phi)$ gives $\pi(U)T(U\to U')=q(U)q(U')\min[r(\Phi),r(\Phi')]$, which is symmetric, so detailed balance holds. One step of the test also keeps a $\Phi$ marginal equal to $p_\Phi$ unchanged, whatever the conditional part is: the error it injects lives only in the modes outside $\Phi$.

\subsection{Overlap and reweighting in the free field}\label{app:free}\label{app:mode}\label{app:corrvol}\label{app:three}

\begin{figure}[t]
\centering
\includegraphics{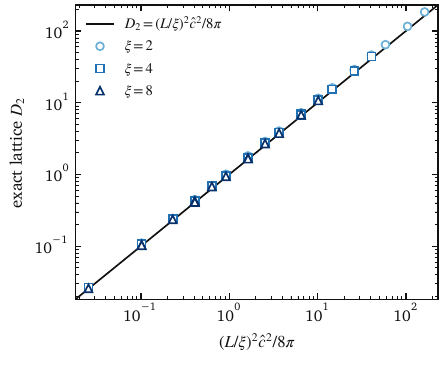}
\caption{Correlation volume law. Exact lattice $D_2$ for the mass mismatch against $(L/\xi)^2\chat^2/8\pi$, for $L=32$--$256$, $L\ge8\xi$, $\chat=0.1$--$0.5$.}
\label{fig:collapse}
\end{figure}

\paragraph{Single Gaussian mode.} For $p\propto e^{-\lambda x^2/2}$ and $q\propto e^{-\lambda(1+\varepsilon)x^2/2}$,
\begin{align*}
\int dx\frac{p^2}{q}&=\frac{\lambda}{2\pi}\Big(\frac{2\pi}{\lambda(1+\varepsilon)}\Big)^{1/2}\int dx\,e^{-\lambda(1-\varepsilon)x^2/2}\\
&=\frac{\lambda}{2\pi}\Big(\frac{2\pi}{\lambda(1+\varepsilon)}\Big)^{1/2}\Big(\frac{2\pi}{\lambda(1-\varepsilon)}\Big)^{1/2}\\
&=\frac{1}{\sqrt{1-\varepsilon^2}},
\end{align*}
for $|\varepsilon|<1$; for $\varepsilon\ge1$ the integral diverges. Independent modes add. A sampling check for the relevant mismatch with $L=32$, $\xi=4$ and $\chat=0.3$ gives $\Eq[w^2]=1.2733(8)$ from $N=10^7$ samples, against $e^{D_2}=1.27338$.

\paragraph{Correlation-volume law and overlap flow.} With $\sum_k\to L^2\int d^2k/(2\pi)^2$ and $u=k^2$,
\begin{align*}
&\frac{\chat^2L^2}{2}\int_{|k|<\Lambda}\frac{d^2k}{(2\pi)^2}\frac{m^4}{(k^2+m^2)^2}\\
&\quad=\frac{\chat^2L^2m^4}{8\pi}\int_0^{\Lambda^2}\frac{du}{(u+m^2)^2}
=\frac{\chat^2L^2m^2}{8\pi}\,\frac{\Lambda^2}{\Lambda^2+m^2}.
\end{align*}
$\Lambda\to\infty$ gives Eq.~\eqref{eq:corrvol}; the ratio gives Eq.~\eqref{eq:flowrel}. At the IR end only the zero mode remains, $-\frac12\log(1-\chat^2)\simeq\chat^2/2$. For the irrelevant case with $m\ll\Lambda\ll1$, $\varepsilon\simeq c_6k^2$ and the same integral gives $L^2c_6^2\Lambda^6/(24\pi)$. For $L=128$, $\xi=4$, $\chat=0.48$ the lattice sum is $10.09$ against $9.39$ from Eq.~\eqref{eq:corrvol}.

Figure~\ref{fig:collapse} compares Eq.~\eqref{eq:corrvol} with the exact lattice sums for $L=32$--$256$, $\xi=2,4,8$ ($L\ge8\xi$) and $\chat=0.1$--$0.5$.  For all 55 points the ratio of the exact sum to Eq.~\eqref{eq:corrvol} lies between $1.01$ and $1.13$; its median is $1.10$, $1.05$, $1.03$ for $\xi=2,4,8$, so  the deviation comes mainly from the lattice dispersion $\kh^2<k^2$, which disappears as $\xi$ grows, plus the $O(\chat^4)$ term in Eq.~\eqref{eq:D2free}. For the one  combination  with $L=4\xi$ in this range ($L=32$, $\xi=8$, not shown) the ratio is $1.27$--$1.39$  for $\chat=0.1$--$0.5$ , because the discrete zero mode is no longer negligible: alone it gives $\chat^2/2$, which is $79\%$ of Eq.~\eqref{eq:corrvol} at $L=4\xi$.

\paragraph{Reweighting estimators.} (i) By independence, \[\Eq[w_AO]=\Eq[w_AO]\,\Eq[w_{\bar A}]=\Eq[wO]=\Ep[O],\] using $\Eq[w_{\bar A}]=1$. (ii) The variance of $x_k$ is $1/[\lambda_k(1+\varepsilon_k)]$ under $q$ and $1/\lambda_k$ under $p$. (iii) $\Eq[w^2(O-\mu)^2]=\Eq[w_{\bar A}^2]\Eq[w_A^2(O-\mu)^2]$ and $\Eq[w_{\bar A}^2]=e^{D_2-D_2(A)}$ by additivity. With interactions, $\Eq[w|\phi_A]=p_A/q_A$ and Jensen's inequality $\Eq[w^2|\phi_A]\ge(p_A/q_A)^2$ show that global reweighting has no smaller asymptotic variance for observables of $\phi_A$, provided the relevant second moments are finite.

\subsection{Composite chain in the free field}\label{app:exact}
(i) The acceptance decision depends only on $x_0$ and the proposed $x_0'$, and the flow \eqref{eq:rotation} of $x_0$ involves only $x_0$, $\pi_0$ and $t$. So $x_0$ alone is a Markov chain. Its first step (the marginal test) preserves $p_0$ by Appendix~\ref{app:stat} with $\Phi=x_0$, and its second step (exact HMC for one Gaussian variable) preserves $p_0$ as well. Hence $p_0$ is the stationary distribution of $x_0$.

(ii) Fix $k\neq0$, and fix the whole history of acceptance decisions and trajectory lengths. This history is independent of the variables of mode $k$ (its proposals $x_k'$ and momenta $\pi_k$): the decisions use only zero mode variables, and the $t_i$ are drawn independently. Given the history, $x_k$ is Gaussian with mean zero; let $v$ be its variance, $v^p=1/\lambda_k$ and $v^q=1/[\lambda_k(1+\varepsilon_k)]$. An accepted proposal sets $v=v^q$; a rejected one leaves $v$ unchanged. By Eq.~\eqref{eq:rotation}, with $\langle\pi_k^2\rangle=1$ and $\pi_k$ independent of $x_k$, one trajectory gives
\begin{align*}
v\ \to\ &\cos^2(\omega_kt)\,v+\frac{\sin^2(\omega_kt)}{\omega_k^2}\\
&=\cos^2(\omega_kt)\,v+\big(1-\cos^2(\omega_kt)\big)\,v^p,
\end{align*}
that is, $v-v^p\to\cos^2(\omega_kt)\,(v-v^p)$.
Starting from $v=v^q$ at the last accepted proposal, $v-v^p=B_k\,(v^q-v^p)$. Averaging over the history, $\mathbb{E}[x_k^2]=v^p+(v^q-v^p)\,\mathbb{E}[B_k]$, and dividing by $v^p$ gives Eq.~\eqref{eq:survival} with $v^q/v^p-1=-\varepsilon_k/(1+\varepsilon_k)$.

(iii) Observations are made at the end of a cycle. The $\nref$ trajectories of that cycle come after the last acceptance decision, so $B_k$ always contains their product $C_k=\prod_{\rm last\ cycle}\cos^2(\omega_kt_i)$, and $B_k\le C_k$ because every factor is at most one. The trajectory lengths of the last cycle are drawn after all decisions, so $\mathbb{E}[C_k]=\rho_k$. This gives the upper bound. If the proposal of the last cycle was accepted, $B_k=C_k$; this event has probability $\bar a$ and is independent of the $t_i$ of that cycle, so $\mathbb{E}[B_k]\ge\bar a\rho_k$.

(iv) For $P(t)=e^{-t/\bar t}/\bar t$,
\[
\mathbb{E}[\cos(2\omega t)]=\frac1{\bar t}\int_0^\infty dt\,e^{-t/\bar t}\cos(2\omega t)=\frac{1}{1+4\omega^2\bar t^2},
\]
so $\mathbb{E}[\cos^2(\omega t)]=\frac12\big(1+\mathbb{E}[\cos(2\omega t)]\big)=\frac{1+2\omega^2\bar t^2}{1+4\omega^2\bar t^2}$. With $u=\omega^2\bar t^2$, $\frac{d}{du}\frac{1+2u}{1+4u}=-\frac{2}{(1+4u)^2}<0$, so this decreases with $\omega$, and $\rho_k\le\rho_m=e^{-\nref/\tau_m}$ for all $k$ because $\omega_k^2=\kh^2+m^2\ge m^2$. Combined with (ii) and (iii) this gives Eq.~\eqref{eq:taum}. For small $u$, $\log[(1+4u)/(1+2u)]\simeq2u$, which gives $\tau_m\simeq\xi^2/(2\bar t^2)$.

\section{Composite chain at fixed correlation length}\label{app:fixedxi}

At $\kappa=0.2705$ the correlation length grows with the box (Sec.~\ref{sec:phi4-setup}), so the runs of Sec.~\ref{sec:phi4chain} do not separate the dependence on $\xi$ from that on $L$. Here we keep $\xi$ fixed. At $\kappa=0.264$, in the symmetric phase, the second-moment correlation length is $\xi\simeq5$ and $\chi_0=50.5$, so that $L=16$--$256$ covers $L/\xi\simeq3$--$50$; benchmark values  and the target marginal $p_M$  at this coupling come from ensembles generated with plain HMC and Wolff updates. As proposal we use the reference process~\eqref{eq:qref} with $\mu_{\rm ref}=0$ and $m_{\rm ref}=0.2106$, for which  $1/K_{\rm ref}(0)=1.87\,\chi_0$ : its low modes exceed the benchmark by up to $87\%$ at $k=0$. We also use two trained samplers at $L=32$ and $64$, with the network and protocol of Table~\ref{tab:osq-samplers} and the reference fixed from the benchmark, $\mu_{\rm ref}=0$ and $m_{\rm ref}=-\hat A(0)+1/\chi_0=0.220$. The chain is that of Sec.~\ref{sec:phi4chain}, with the marginal test on $M$; at $L\ge128$ the HMC step is reduced so that the HMC acceptance rate stays near $0.9$ (Table~\ref{tab:tau}).

\begin{figure*}[t]
\centering
\includegraphics[width=\textwidth]{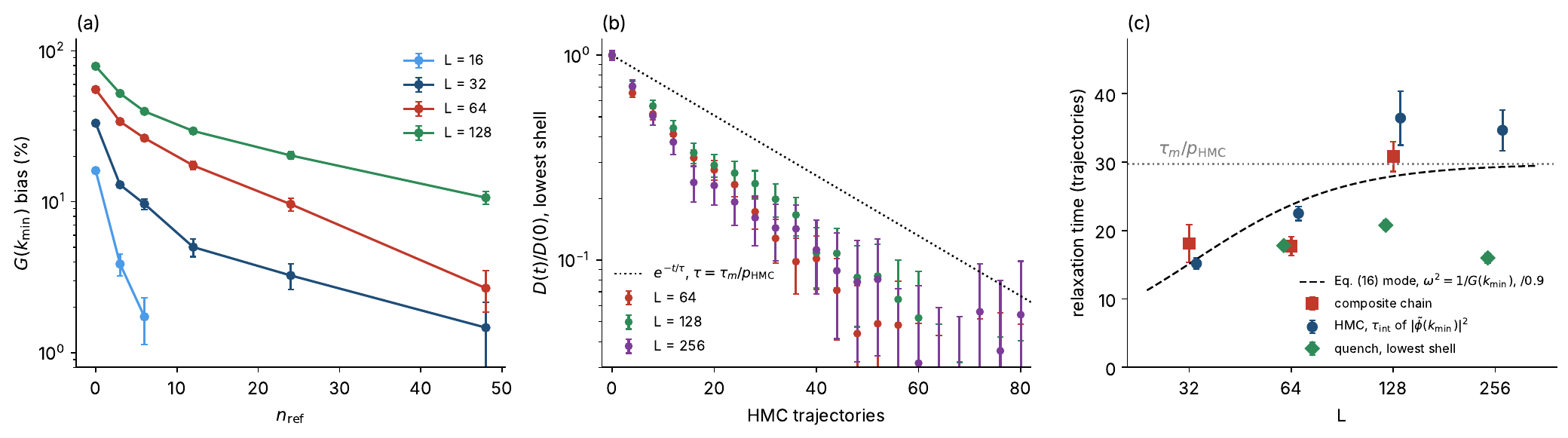}
\caption{Fixed correlation length, $\kappa=0.264$ ($\xi\simeq5$). (a) Bias of $G(k_{\min})$, i.e.\ its relative deviation from the benchmark value, in the composite chain against $\nref$, with the reference process as proposal; points consistent with zero are omitted. (b) Two-sided quench under plain HMC: difference $D(t)$ between  reference-process  starts with too strong and too weak low modes, shell $1\le|n|<2$, normalized to $D(0)$, with $p_{\rm HMC}=0.905$ at all $L$. Dotted: decay with $\tau_m/p_{\rm HMC}$. (c) Relaxation times of Table~\ref{tab:tau} against $L$. Dashed: the relaxation time~\eqref{eq:taum} of a mode with $\omega^2=1/G(k_{\min})$, that is, Eq.~\eqref{eq:tauk}, with $G(k_{\min})$ from Eq.~\eqref{eq:Gkmin} at $\xi=5.3$, divided by $p_{\rm HMC}=0.9$; dotted: the free-field estimate $\tau_m/p_{\rm HMC}$.}
\label{fig:fixedxi}
\end{figure*}

\begin{table*}[t]
\centering\small
\begin{tabular}{ccccccc}
\toprule
$L$ & $\epsilon$ & $p_{\rm HMC}$ & $\tau_k/p_{\rm HMC}$, Eq.~\eqref{eq:tauk} & chain & quench & equilibrium $\tau_{\rm int}$\\
\midrule
32  & 0.1  & 0.95 & 14.5 & 18(3)     & --      & 15.2(8)\\
64  & 0.1  & 0.90 & 24.0 & 17.7(14)  & 17.8(5) & 22.5(10)\\
128 & 0.1  & 0.82 & 30.4 & 30.8(22)  & --      & --\\
128 & 0.07 & 0.90 & 27.4 & 29(3)     & 20.8(7) & 36(4)\\
256 & 0.05 & 0.90 & 28.7 & --        & 16.0(7) & 35(3)\\
\bottomrule
\end{tabular}
\caption{Relaxation time of the lowest nonzero modes at $\kappa=0.264$ ($\xi\simeq5$), in HMC trajectories. Chain: exponential fit of the deviation of $G(k_{\min})$ from the benchmark for $\nref\ge6$, with the reference process as proposal. Quench: exponential fit of $D(t)$ in the shell $1\le|n|<2$. Equilibrium: integrated autocorrelation time of $|\tilde\phi(k_{\min})|^2$ under plain HMC, from 8 chains of 6000 trajectories. The fourth column is Eq.~\eqref{eq:tauk} with the measured $G(k_{\min})$, divided by $p_{\rm HMC}$.}
\label{tab:tau}
\end{table*}

\paragraph{Lowest nonzero mode.} The marginal test alone has the target zero-mode distribution at stationarity when its density ratio is known. Local updates must repair the other low modes and can also disturb the zero mode. Among the untested modes, the lowest, $G(k_{\min})$, is expected to relax most slowly. Figure~\ref{fig:fixedxi}(a) shows its deviation from the benchmark value. At $\nref=0$ it is $+16$, $33$, $55$ and $79\%$ at $L=16$, $32$, $64$ and $128$. These are the errors of the proposal for this mode, $1/K_{\rm ref}(k_{\min})$ against the benchmark $G(k_{\min})$, which are $+16$, $31$, $59$ and $81\%$: they grow with $L$ because $k_{\min}=2\pi/L$ moves into the region $k\lesssim1/\xi$ where the proposal is wrong, and they are bounded by the error at $k=0$. This is the unweighted bias of Eq.~\eqref{eq:survival}.

\paragraph{Relaxation time.} Proposition~\ref{prop:exact}(iv) gives the free-field answer. In the Gaussian approximation, a mode with propagator $G(k)$ has frequency $\omega_k^2=1/G(k)$ in HMC with unit-mass momenta, and its bias decays per trajectory by the factor $\mathbb{E}_t[\cos^2(\omega_kt)]$. With the measured $G(k)$, the free-field formula gives the estimate
\begin{equation}\label{eq:tauk}
\begin{split}
\tau_k&=\Big[\log\frac{1+4/G(k)}{1+2/G(k)}\Big]^{-1}\\
&\le\tau_m=\Big[\log\frac{1+4/\chi_0}{1+2/\chi_0}\Big]^{-1}=26.8,
\end{split}
\end{equation}
where the bound uses $G(k)\le G(0)=\chi_0$ in the symmetric phase (for $\bar t=1$). Since a rejected trajectory leaves the configuration unchanged, the time in trajectories is about $\tau_k/p_{\rm HMC}$, where $p_{\rm HMC}$ is the HMC acceptance rate. How close $k_{\min}$ is to the bound follows from the small-$k$ form $G(k)\simeq\chi_0/(1+\xi^2k^2)$:
\begin{equation}\label{eq:Gkmin}
\begin{split}
\frac{G(k_{\min})}{\chi_0}&\simeq\frac1{1+(2\pi\xi/L)^2}\\
&=0.51,\ 0.80,\ 0.94,\ 0.985\\
&\qquad(L=32,\,64,\,128,\,256),
\end{split}
\end{equation}
against the measured $0.49$, $0.79$ and $0.93$ at $L\le128$. This estimate predicts that the relaxation time grows from $L=32$ to $128$ and then approaches a finite limit.

\begin{figure*}[t]
\centering
\includegraphics[width=\textwidth]{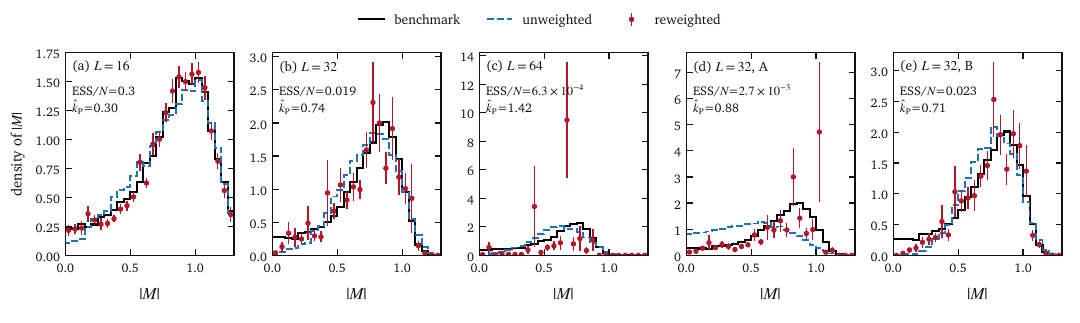}
\caption{ Reweighting the OSQ samples with their path weights. Distribution of $|M|$ from the benchmark ensembles (solid), the unweighted samples (dashed) and the reweighted samples (points, with delta-method errors) for the samplers with fixed reference at (a) $L=16$, (b) $L=32$ and (c) $L=64$, and for samplers (d) A and (e) B at $L=32$. Each panel gives $\ESS/N$ and the Pareto shape $\hat k_{\rm P}$ of the largest weights. }
\label{fig:reweight}
\end{figure*}

We measured the relaxation time in three ways (Table~\ref{tab:tau} and Fig.~\ref{fig:fixedxi}): from exponential fits of the deviation in the chain for $\nref\ge6$; from a two-sided quench under plain HMC, in which configurations of the reference process with  $1/K_{\rm ref}(0)=1.87\,\chi_0$  (low modes too strong) and with $1/K_{\rm ref}(0)=0.5\,\chi_0$ (too weak), in both cases with the zero mode rescaled to variance $\chi_0/V$ as after a marginal test, are evolved and the difference $D(t)$ of the mean of $|\tilde\phi(k)|^2/V$ over the shell $1\le|n|<2$ is followed [Fig.~\ref{fig:fixedxi}(b)]; and from the integrated autocorrelation time of $|\tilde\phi(k_{\min})|^2$ under plain HMC in equilibrium. The quench and equilibrium estimates show no growth between $L=128$ and $256$; the chain result at $L=128$ is compatible with the same relaxation scale. None of the measured times exceeds the free-field reference scale $\tau_m/p_{\rm HMC}\simeq30$ by more than $20\%$ [Fig.~\ref{fig:fixedxi}(c)]. The quench also resolves the momentum dependence: shells at the same physical $k$ relax at similar rates across the volumes studied, in $6.1(3)$, $6.3(3)$ and $5.6(2)$ trajectories at $k\simeq0.3$ for $L=64$, $128$ and $256$, and in $13.8(4)$ and $13.0(3)$ trajectories at $k\simeq0.15$ for $L=128$ and $256$. The  estimators  differ by up to a factor $1.7$  at $L\le128$ and $2.2$ at $L=256$; they need not agree, since  they measure different things: the quench the linear response averaged over a shell, the chain fit the late part of a large one-sided deviation, and $\tau_{\rm int}$ also the slowest components of the equilibrium fluctuations. At fixed $\xi$ these data support saturation of the number of trajectories per cycle once $L$ is several times $2\pi\xi$, as in the free field (Sec.~\ref{sec:exact}), while the cost of one trajectory grows like $V^{5/4}$, as for plain HMC (Sec.~\ref{sec:phi4chain}).

\paragraph{Trained samplers.} An exponential relaxation estimate gives $\nref\simeq\tau\log(b_0/\delta)$ for a residual bias $\delta$, where $b_0$ is the initial bias in the low mode of interest. For the reference process $b_0\simeq0.8$, and about 110 trajectories are needed for $2\%$. The trained samplers reproduce the  low-mode propagator: at the lowest momenta, which include all $k\lesssim1/\xi$, their propagator is within $1$--$4\%$ of the benchmark, and $\chi_0$ within $1.5\%$, against $+87\%$ at $k=0$ for the reference process.  This propagator is essentially that of their own reference, whose $m_{\rm ref}$ is fixed from the benchmark $\chi_0$: its $1/K_{\rm ref}(k_{\min})$ is about $0.49\,\chi_0$ and $0.79\,\chi_0$ at $L=32$ and $64$, as is the benchmark $G(k_{\min})$. The contrast with the reference process thus reflects the choice of $m_{\rm ref}$ rather than training.  In the chain, $G(k_{\min})$ deviates by $+4.4(17)\%$ at $L=32$ and $+2.4(17)\%$ at $L=64$ at $\nref=0$, and is consistent with zero within two standard deviations from $\nref=6$ on; the marginal test accepts $0.96$--$0.99$ of the proposals, and $\tau_{\rm int}(M^2)$ is $0.5$ cycles. With $b_0\simeq0.03$ a few trajectories suffice. The reference process in the same setup, run with the second implementation, starts from $+28(3)\%$ and $+62(3)\%$ and pulls the zero mode as at $\kappa=0.2705$: $\chi_0$ dips by $11\%$ and $15\%$ at $\nref=6$, whereas for the trained samplers it stays consistent with zero. At fixed $\xi$, too, the generator has to reproduce $G(k)$ for $k\lesssim1/\xi$ ; the two trained samplers do so, and  the local updates have little left to  correct in $G(k_{\min})$ and $\chi_0$ .

\section{\texorpdfstring{ Estimating and using the target marginal }{Estimating and using the target marginal}}\label{app:pphi}

 The marginal test needs the target marginal $p_\Phi$ as input (Sec.~\ref{sec:fix}). For $\Phi=M$ at $\kappa=0.2705$, we compare two ways of estimating $p_M$: reweighting the samplers' own configurations and constrained simulations of the action. We then use the constrained estimate to construct proposals with a prescribed distribution of $M$, and explore the effect of supplying additional low modes from the benchmark ensembles. 

\begin{figure*}[t]
\centering
\includegraphics{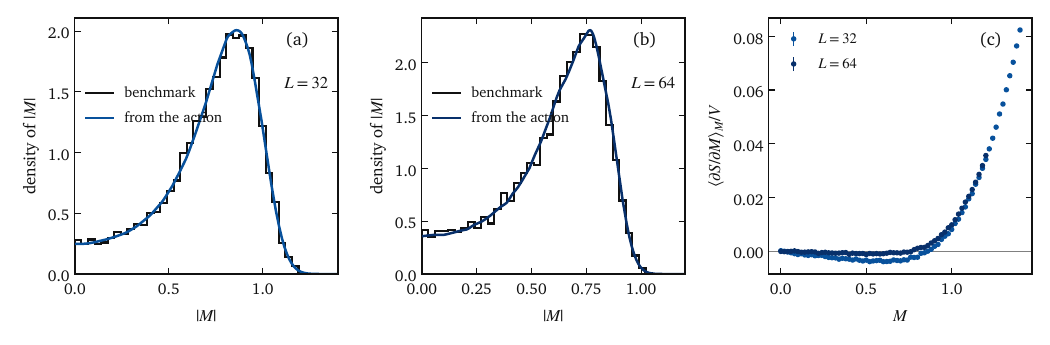}
\caption{ Distribution of the magnetization from the action at $\kappa=0.2705$. (a), (b) Density of $|M|$ from the benchmark ensembles (histogram) and $\hat p_M$ obtained by integrating the mean force at fixed $M$ (curve), at $L=32$ and $64$. (c) Mean force $\langle\partial S/\partial M\rangle_M/V$ against $M$, with jackknife errors. Each fixed-$M$ HMC chain starts from an OSQ sample shifted to the prescribed $M$; starts from the reference process give consistent results within errors. }
\label{fig:ti}
\end{figure*}

\paragraph{ Reweighting with the path weights. }
 The exact OSQ path weights recover the target distribution of $M$ asymptotically, but with $N=10240$ samples this works only at the smallest volume (Fig.~\ref{fig:reweight}). At $L=16$, where $\ESS/N=0.30$, $\langle|M|\rangle$, $\chi_0$, $\chi$ and $U_L$ agree with the benchmark within $1.5$ standard deviations. At $L=32$ and $64$ the Pareto shapes of the largest weights, $\hat k_{\rm P}=0.74$ and $1.42$, exceed the reliability threshold $0.7$~\cite{Vehtari:2024}; at $L=64$, three configurations carry half of the total weight. Poor coverage also matters: sampler B has almost no configurations at small $|M|$, and its reweighted $\chi$ remains $28(5)\%$ low. Thus, at the available sample size, the samplers' own weights do not reliably determine $p_M$ at the larger volumes. 

\paragraph{ Constrained simulations. }
 An alternative is to hold $M$ fixed while HMC relaxes the remaining modes. The marginal follows by integrating the mean force, as in the Monte Carlo determination of the constraint effective potential~\cite{ORaifeartaigh:1986}: 
\begin{equation*}
 \begin{aligned}
-\partial_M\log p_M(M)&=\left\langle\frac{\partial S}{\partial M}\right\rangle_M\\
&=V\hat A(0)M+4\lambda\left\langle\sum_x\phi_x^3\right\rangle_M.
\end{aligned} 
\end{equation*}
 We used a grid of spacing $0.02$ on $0\le M\le1.40$ at $L=32$ and $0\le M\le1.20$ at $L=64$, with a total of 64 and 32 chains per grid point across four independent runs. Each chain started from an OSQ sample, reflected if necessary and shifted uniformly to the grid value. We discarded the first 80 of 300 and 400 trajectories per chain, respectively. 

 The reconstructed $\hat p_M$ agrees with the benchmark (Fig.~\ref{fig:ti}): $\langle|M|\rangle$, $\chi_0$ and $U_L$ deviate by less than $1\%$, and $\chi$ by $-1.0(19)\%$ and $-2.1(29)\%$ at $L=32$ and $64$, all within one standard deviation. Errors are the largest of the jackknife errors over chains and over groups of chains and the scatter of the four runs. Gaussian reference starts give consistent results; at $L=64$, discarding 200 instead of 80 trajectories changes $\chi$ by $0.2\%$ for the OSQ starts. Thermalization is nevertheless necessary: without discarding trajectories, $\chi$ is $9(3)\%$ low for the reference starts at $L=64$. Both initializations use benchmark-calibrated $m_{\rm ref}$ and $\mu_{\rm ref}$ (Table~\ref{tab:osq-samplers}); starts without this information have not been tested. 

 The calculation used $1.4\times10^6$ and $7.8\times10^5$ HMC trajectories, including those discarded. For the same statistical error in $\chi$, plain HMC started from equilibrium configurations is estimated to require 1.6--2.3 times as many trajectories at both volumes. This comparison uses 16 plain HMC chains per volume and the conservative errors of $\hat p_M$ above; the relevant autocorrelation time, that of $V(M^2-2\langle|M|\rangle|M|)$, is 109 and 345 trajectories. The trajectory counts do not include training or generation, and no gain growing with volume is resolved. 

\paragraph{ Proposals with a known marginal. }
 Given $\hat p_M$, we draw $M$ directly, take an unused OSQ sample with the nearest $|M|$, reflect it if necessary, shift it uniformly to the drawn value, and apply 60 HMC trajectories at fixed $M$. These proposals have marginal $\hat p_M$, so the marginal test with this target accepts all of them and no estimate of $q_M$ is needed. We then apply $\nref$ plain HMC trajectories. At $\nref=0$, observables of $M$ alone follow from the input marginal and provide no independent validation; the other observables probe the conditional distribution. 

 After the 60 fixed-$M$ trajectories, $\langle\phi_x^2\rangle$ and $\langle S\rangle/V$ agree with the benchmark at both volumes within statistical errors of at most $0.3\%$, and $G(k_{\min})$ is consistent with the benchmark at $L=32$. At $L=64$, however, $G(k_{\min})$ remains $10(3)\%$ low and does not change significantly during 40 further unconstrained trajectories, as in the composite chain [Fig.~\ref{fig:chainosq}(b)]. This mode must be better reproduced by the sampler, relaxed longer, or included in $\Phi$. 

 The construction can also start from the Gaussian reference process instead of a trained generator. With the same 60 fixed-$M$ trajectories, these observables are again consistent with the benchmark at $L=32$. At $L=64$, $G(k_{\min})$ is $22(2)\%$ low, and the local observables $\langle\phi_x^2\rangle$ and $\langle S\rangle/V$ also deviate, by $-0.9(3)\%$ and $+0.45(11)\%$. The trained sampler thus leaves smaller residual deviations at the same relaxation length; this comparison does not establish a net computational advantage including training and generation. 

\paragraph{ Supplying the slow modes. }
 At $L=64$, longer relaxation at fixed $M$ reduces the discrepancy in $G(k_{\min})$ only slowly: after 120 and 240 trajectories it is $6(3)\%$ and $4(3)\%$ low. Rescaling the nonzero Fourier amplitudes shell by shell for $\kh<0.6$ matches the propagator of the OSQ pool to the benchmark, but does not ensure equilibrium: selecting samples by $M$ can change these second moments, and subsequent relaxation lowers $G(k_{\min})$ again. After 60 fixed-$M$ trajectories it is $9(3)\%$ low, as without rescaling. Holding the rescaled modes fixed postpones the change until release. Propagator calibration is therefore insufficient; the failure is consistent with, but does not by itself establish, an error in the joint distribution of the slow modes. 

 To see whether supplying the joint distribution of the low modes, not only their propagator, prevents the decrease of $G(k_{\min})$ under subsequent relaxation, we next combined the 121 real Fourier degrees of freedom with $\kh<0.6$, including the zero mode, from each benchmark configuration with the remaining modes from an OSQ sample. The mixed samples were selected, reflected and shifted to $M$ drawn from $\hat p_M$ as above; the joint infrared input inherits the errors of $\hat p_M$ and finite-pool matching. HMC with these low modes held fixed targets the conditional distribution of the remaining modes. At this stage $G(k_{\min})$ is preserved by construction, so its agreement with the benchmark is not a thermalization test. 

 During 10 such constrained trajectories the action density relaxes: its excess drops from $2.7\%$ to $0.01(10)\%$, and $\langle\phi_x^2\rangle$ agrees with the benchmark within errors. After the low modes are released, no drift is resolved during 40 further unconstrained trajectories: both local observables remain consistent with the benchmark, and $G(k_{\min})$ deviates by $-0.3(31)\%$, that is, it is compatible with the benchmark at a statistical precision of about $3\%$. With only the 29 real degrees of freedom below $\kh=0.3$, 10 constrained trajectories leave an action-density excess of $0.23(10)\%$; after 10 additional unconstrained trajectories it is $0.06(10)\%$, and the measured local observables and $G(k_{\min})$ remain consistent with the benchmark through 40 unconstrained trajectories. The benchmark supplies both the low-mode input and the comparison values, so this tests conditional relaxation rather than independently validating the full distribution. 

 The $0.6$-cutoff test uses one OSQ sample and 10 constrained trajectories; the $0.3$-cutoff test also needs 10 unconstrained trajectories for the reported agreement. These counts per output configuration cannot be compared directly with the plain-HMC cost per effective independent $M^2$ sample in Table~\ref{tab:cost}. An efficiency comparison at equal accuracy must also include the infrared input, training, and the overhead of Fourier projections. An RG-matched coarse ensemble~\cite{Hasenfratz:2026,Hasenfratz:2026b} could supply such infrared information. A marginal test on $M$ could then correct its magnetization marginal if the required density ratio were known, while errors in the other low modes and their correlations could remain. This hybrid has not been tested. 

\bibliography{references}

\end{document}